\documentclass[11pt,a4paper]{JHEP3}
\usepackage{amsmath, amssymb}
\usepackage{euscript}
\def\be{\begin{eqnarray}}
\def\ee{\end{eqnarray}}
\def\0{\nonumber}
\def\d{\partial}

\newcommand\ET{\EuScript{T}}
\usepackage{amsfonts}
\usepackage{verbatim}
\newcommand{\CR}{\\\nonumber}

\newcommand{\refb}[1]{(\ref{#1})}
\newcommand{\NO}[1]{\ensuremath{\,:\!#1\!:\,}}
\newcommand{\VEV}[1]{\ensuremath{\,\langle{}#1\rangle\,}}

\newcommand{\Tr}{\mathrm{Tr}}

\def\k{\kappa}
\def\l{\lambda}
\def\u{\tilde u}

\preprint{SISSA/50/2008/EP\\\tt hep-th/0808.2360}

\title{Hawking fluxes, $W_\infty$ algebra and anomalies}

\author{ L.Bonora$^a$, M.Cvitan$^{a,b}$, S.Pallua $^{b}$, 
I.Smoli\'c $^b$\\
 $^a$ International School for Advanced Studies (SISSA/ISAS)\\
Via Beirut 2--4, 34014 Trieste, Italy, and INFN, Sezione di
Trieste\\
  $^b$ Theoretical Physics Department, Faculty of Science,
        University of Zagreb\\
        p.p. 331, HR-10002 Zagreb, Croatia\\

E-mail:   \email{bonora@sissa.it}, \email{cvitan@sissa.it}}

\abstract{We complete the analysis started in [arXiv:0804.0198] of 
the Hawking radiation calculated by means of anomaly techniques.
We concentrate on a static radially symmetric BH, reduced to two 
dimensions. We compare the two methods used to derive the integrated
Hawking radiation, based on the trace and diffeomorphism anomaly, 
respectively, and show that they can be reduced to the 
same basic elements. We then concentrate on higher moments of the Hawking 
radiation and on higher spin currents, and show that, similarly to
trace anomalies, also diffeomorphism anomalies are absent from the
conservation laws of higher spin currents. We show that the
predictivity of the method is due to the $W_\infty$ current algebra
underlying the effective model that describes matter around the black 
hole.}


\begin{document}

\maketitle

\section{Introduction}

Hawking radiation \cite{Hawking1,Hawking2} is a universal phenomenon
which does not depend on the details of the collapse that gives
rise to a black hole. Therefore one would expect that there exist methods
to calculate it that have the same character of universality. Local anomalies 
have such a characteristic, because all anomalies have a universal form,
only the coefficients in front of them are model dependent. A first attempt
to compute Hawking radiation by exploiting trace anomalies was made long time ago
by Christensen and Fulling, \cite{CF}, and reproposed subsequently
by \cite{Thorlacius,Strominger} in a modified form. More recently
a renewed attention to the same problem has been pioneered by the paper 
\cite{Robinson}, where diffeomorphisms anomalies have been used instead 
of trace anomalies. This paper is at the origin of a considerable activity
with numerous contributions \cite{IUW1,IUW2,IMU1,IMU2,IMU3,
IMU4,IMU5,
Murata:2006pt,
Vagenas:2006qb,
Setare:2006hq,
Jiang:2007gc,
Jiang:2007pn,
Jiang:2007wj,
Kui:2007dy,
Shin:2007gz,
Jiang:2007mi,
Das:2007ru,
Chen:2007pp,
Miyamoto:2007ue,
Jiang:2007pe,
Kim:2007ge,
Murata:2007zr,
Peng:2007nj,
Ma:2007xr,
Huang:2007ed,
Peng:2008ru,
Wu:2008yx,
Gangopadhyay:2008zw,
Kim:2008hm,
Xu,Banerjee1,Banerjee2,Gango,Gango2,Kulkarni,Peng1,Peng2,Peng3,Iso:2008sq,Umetsu:2008cm,
Shirasaka:2008yg,Ghosh:2008tg,Banerjee:2008ez,Banerjee:2008az}. 

The purpose of the present paper, which is a sequel to \cite{BC}, is to assess the role of 
anomalies in computing the thermal spectrum of the Hawking radiation. Our  
conclusion is that, while anomalies (trace or diffeomorphism) can be used to
compute the integrated Hawking radiation, this is not the case for 
higher moments. Rather we find that there exists an underlying structure at the basis
of the universality of Hawking radiation: this is a $W_\infty$ algebra which
characterizes the underlying matter model describing the radiation.

In this paper, as in \cite{BC}, in order to be able to discuss the essential aspects
while avoiding inessential complications, we will stick to the simplest case of a 
static chargeless black hole with metric
\be
ds^2= f(r) dt^2 - \frac 1{f(r)}dr^2 -r^2 d\Omega^2\label{4metric}
\ee
$f(r)$ near the horizon behaves like $f(r)\approx 2 \k (r-r_H)$, where 
$\k$ is the surface gravity. An essential step in this kind of approach is the
reduction to a two--dimensional problem. This can be done by using 
radial symmetry, postulating the independence of the polar coordinates
$\theta,\varphi$ and expanding the fields in spherical harmonics.
For instance, for a scalar field, $\phi(t,r,\theta,\varphi)=\sum_{lm}
Y_{lm}(\theta,\varphi)\,\phi_{lm}(t,r)$. One then integrates,  
in the action, over the polar angles. This has been done in some details,
for instance, in \cite{IMU1}, the result being a theory of infinite many
complex scalar fields $\phi_{lm}$ interacting with the background gravity
specified by the metric 
\be
ds^2= f(r) dt^2 - \frac 1{f(r)}dr^2 \label{2metric}
\ee
In the following we will retain only one of all these complex scalar 
fields. The analysis for all the other scalar fields is the same, what 
is left out from our analysis is how to resum all these contributions
and obtain some four--dimensional information (see however the comment 
at the end of section 3). 
 
In the first part of our paper we review the two methods based on anomalies, 
the diff and trace anomaly method. The purpose is to stress that they are actually
based on the same basic formulas and same basic requirements (no ingoing flux from
infinity and vanishing of energy--momentum tensor at the horizon).
Next we take up the problem of higher moments of the Hawking radiation.
Following \cite{IMU1,IMU2,IMU3,IMU4}, we attribute these higher fluxes to 
phenomenological higher spin currents, i.e. higher spin generalizations of the 
energy--momentum tensor. In \cite{BC} it was shown that these currents can be 
constructed out of a $W_\infty$ algebra. It is the properties of this $W_\infty$ algebra
that explain the higher moments of Hawking radiation.
As was shown in \cite{BC} the higher spin currents are not anomalous, at variance 
with \cite{IMU3,IMU4}, where, in a different (spinorial) matter model, anomalies 
were found in the conservation laws and traces of higher spin currents. In this
paper we complete the analysis started in \cite{BC}, where, using consistency methods, 
the absence of true trace anomalies was proved at least for the fourth order 
current. Here we deal with the far more complicated case of diff anomalies. The result
is invariant: there cannot exist any true diff anomalies in the fourth order current.
This confirms a well founded prejudice according to which true gravitational anomalies can
exist when there is a precise correspondence between number of derivative in the
anomaly polynomial and space--time dimensions.

The conclusion of our analysis is that the universal element that explains the 
universal character of the Hawking fluxes lies in the $W_\infty$ algebra 
underlying the matter model for radiation.

\section{Review of the anomaly methods}

In \cite{Robinson} the method used was based on the 
diffeomorphism anomaly in a two--dimensional effective field theory
near the horizon of a radially symmetric static black hole.
The basic argument is that, since just outside the horizon the ingoing modes
cannot classically influence the physics outside the black hole, they can be
integrated out, giving rise to an effective theory of purely outgoing 
modes. So the physics in that region can be described by 
an effective two--dimensional chiral field theory
(of infinite many fields). This implies 
an effective breakdown of the diffeomorphism invariance. The ensuing 
anomaly equation can be utilized to compute the outgoing flux of radiation. 
The latter appears as the quantum factor that restores the 
diffeomorphism symmetry. 

\subsection{Diff anomaly method}

Let us describe in detail the corresponding derivation 
as given, in a somewhat simplified form, in \cite{Banerjee2}
\footnote{After completion of this paper one of the author, R.Banerjee, 
has pointed out to us that the diff anomaly 
method can be further simplified by using a single Ward identity
instead of two as in the presentation below. This does not change
however our conclusions in section 3.}.
The range of $r$ contains two relevant regions: the region $o$, 
defined by $r> r_H+\epsilon$, $r_H$ being the horizon radius, and the 
region $H$, defined by $r_H< r< r_H+ \epsilon$.
The region $H$ is where the ingoing modes have been integrated out, therefore
the effective field theory there is anomalous, while in $o$ we expect a fully
symmetric theory. This is expressed by a vanishing energy momentum tensor
covariant divergence
\be
\nabla_\mu T^\mu {}_{\nu(O)}=0,\label{div0}
\ee
while in the $H$ region we have 
\be
\nabla_\mu T^\mu {}_{\nu(H)}=  \frac{\hbar c_R}{96 \pi} \epsilon_{\nu\mu} \partial^\mu R
\label{diffanom}
\ee
This is the covariant form of the diffeomorphism anomaly, with a coefficient 
appropriate for chiral (outgoing or right) matter with central charge $c_R$. 
In \refb{diffanom}
$\epsilon_{\mu\nu}= \sqrt{-g} \varepsilon_{\mu\nu}$, where $ \varepsilon$ is the
numerical antisymmetric symbol ($\varepsilon_{01}=1$). In the case of the
background metric we are considering, the determinant is -1. Since the metric is also
static, the two equations above take, for $T_t^r$, a very simple form:
\be
\d_rT^r_{t(o)}=0\label{div01}
\ee
and 
\be
\d_rT^r_{t(H)}=\d_r N^r_t\equiv \d_r\left(\frac {\hbar c_R}{96\pi}(f f''-
 \frac 12 (f')^2)\right)
\label{diffanom1}
\ee 
respectively.
Now we integrate these equations in the respective regions of validity
\be
T^r_{t(o)}= a_o\label{div02}
\ee
and 
\be
T^r_{t(H)}(r)= a_H + N^r_t(r)-N^r_t(r_H)\label{diffanom2}
\ee
We remark that $a_o$, being constant, determines (together with the 
condition
that there is no ingoing flux from infinity) the outgoing energy flux.
This is the quantity we would like to know. To this end we define 
the overall energy--momentum tensor.
\be
T^r_t= T^r_{t(o)} \theta(r-r_H-\epsilon)+ T^r_{t(H)} 
\left(1-\theta(r-r_H-\epsilon){}^{}\right)
\label{overall}
\ee
It is understood that $\epsilon$ is a small number which 
specifies the size of the 
region where the energy--momentum tensor is not conserved. 
If we take the divergence of \refb{overall}, we get
\be
\d_r T^r_t =\left(a_o-a_H+N_t^r(r_H){}^{}\right)\delta(r-r_h-\epsilon)
+\d_r \left(N^r_t(r){}^{} H(r)\right)
\ee
where $H(r)= 1-\theta(r-r_H-\epsilon)$. We can now define a new overall 
tensor
\be
\hat T^r_t(r)= T^r_t(r)-N^r_t(r) H(r)\label{overallem}
\ee
which is conserved
\be
\d_r \hat T^r_t=0\label{newcons}
\ee
provided that 
\be
a_o-a_H+N_t^r(r_H)=0\label{conscond}
\ee
Now, the condition that at the horizon the energy--momentum tensor vanishes,
leads to $a_H=0$ (see \refb{diffanom2}). Therefore
\be
a_o=N^r_t(r_H)= \frac {\hbar\k^2 }{48\pi}c_R\label{rad}
\ee
This is the outgoing flux at infinity and coincides with the total Hawking
radiation (see below) emitted by the black hole specified by the metric 
\refb{2metric}. We remark that $\hat T^r_t$ is constant everywhere.

\subsection{Trace anomaly method}

The method based on the trace anomaly was suggested long ago
by Christensen and Fulling, \cite{CF} (see also \cite{Davies}). Such a method 
has been reproposed in different forms in \cite{Thorlacius,Strominger} 
and, in particular, \cite{IMU2} and \cite{IMU4} (see also \cite{BC}).  
This approach is based on the argument that the near--horizon physics is described by a 
two--dimensional conformal field theory (see also \cite{Carlip,Solod,Balbinot}). 
Classically the trace of the matter energy momentum tensor vanishes on shell.
However it is generally nonvanishing at one 
loop, due to the anomaly: $T_\alpha^\alpha= \frac {c}{48\pi} R$, where 
$R$ is the background Ricci scalar. $c$ is the total central charge of the 
matter system. The idea is to use this piece of information in order to compute
the same constant $a_o$ calculated with the previous method. Here we do not have 
to split the space in different regions, but we consider 
a unique region outside the horizon.

With reference to the metric \refb{2metric} it is convenient to transform it
into a conformal metric. This is done by means of the 'tortoise' coordinate $r_*$ 
defined via $\frac {\partial r}{\partial r_*}=f(r)$. Next 
it is useful to introduce light--cone coordinates $u=t-r_*, v=t+r_*$. Let us denote
by $T_{uu}(u,v)$ and $T_{vv}(u,v)$ the classically non vanishing
components of the energy--momentum tensor in these new coordinates. 
Our black hole is now characterized by the background metric 
$g_{\alpha\beta} = e^\varphi \eta_{\alpha\beta}$, where $\varphi= \log f$.
The energy--momentum tensor can be calculated by integrating the conservation
equation and using the trace anomaly. The result is (see next section)
\be
T_{uu}(u,v)= \frac {\hbar c_R} {24\pi} \left(\d_u^2 \varphi- 
\frac 12 (\d_u\varphi)^2\right)+T^{(hol)}_{uu} (u)\label{TuuThol}
\ee
where $T_{uu}^{(hol)}$ is holomorphic, while $T_{uu}$ is conformally 
covariant. Namely, under a conformal transformation $u\to \tilde u=f(u)
(v\to \tilde v=g(v))$ one has
\be
T_{uu}(u,v)=\left(\frac {df}{du}\right)^2 T_{\tilde u\tilde u}
(\tilde u, v)\label{conf}  
\ee 
Since, under a conformal transformation, $\tilde \varphi(\u,\tilde v)=
\varphi(u,v) -\ln \left( \frac {df}{du}\frac {dg}{dv}\right)$, it follows 
that 
\be
T^{(hol)}_{\u \u}(\u) = \left(\frac {df}{du} \right)^{-2} \left(
T^{(hol)}_{u u}(u)+ \frac {\hbar c_R}{24\pi} \{\u,u\}\right)\label{Thol}
\ee
Regular coordinates near the horizon are the Kruskal ones, $(U,V)$, 
defined by
$U=-e^{-\k u}$ and $V=e^{\k v}$. Under this transformation we have
\be
T^{(hol)}_{UU}(U) = \left(\frac 1{\k U} \right)^{2} \left(
T^{(hol)}_{u u}(u)+ \frac {\hbar c_R}{24\pi} \{U,u\}\right)\label{TholU}
\ee
Now we require the outgoing energy flux to be regular at the future
horizon $U=0$ in the Kruskal coordinate. Therefore at that point 
$T^{(hol)}_{u u}(u)$ is given by $\frac  {c_R \k^2}{48 \pi}$.
We remark that this implies in particular that
$T_{uu}(r=r_H)=0$.
 
Since the background is static, $T^{(hol)}_{u u}(u)$
is constant in $t$ and therefore also in $r$. Therefore at $r=\infty$
it takes the same value $\frac {\hbar c_R \k^2}{48 \pi} $.
On the other hand we can assume that at $r=\infty$ 
there is no incoming flux and that the background is 
trivial  (so that the vev of $T^{(hol)}_{u u}(u)$ and $T_{u u}(u,v)$ 
asymptotically coincide)\footnote{We stress that vanishing of $\VEV{T_{vv}}$ 
does not contradict the stress tensor conservation. $T_{vv}$ has an expression
similar to \refb{TuuThol}, with subscripts $u$ replaced by $v$ and $c_R$ replaced by
$c_L$, see \refb{Tvv} below. Since $T_{vv}^{a-hol}$ vanishes at infinity and is 
conserved, it would seem at first that this leads to a contradiction with 
a formula similar to \refb{TholU} for the ingoing part. We notice however
that $V=0$ is not the future horizon and no vanishing condition for the 
stress tensor is required there.}.

Therefore the asymptotic flux is 
\be
\langle T_t^r\rangle =\VEV {T_{uu}}-\VEV {T_{vv}} = 
\frac{\hbar \k^2}{48 \pi} c_R\label{flux}
\ee
This outgoing flux coincides with the constant $a_o$ calculated above.

In summary we can say that the basic ingredients of the two methods are:

\begin{itemize}
\item (a) in the first case the integration of the anomalous and non-anomalous 
conservation of the energy-momentum tensor, in the second case the
integration of the energy--momentum conservation in the presence of a 
trace anomaly; 

\item (b) in both cases we have the condition that the energy--momentum
tensor vanishes at the horizon and there is no incoming energy flux 
from infinity.
\end{itemize}

What energy--momentum tensor vanishes at the horizon will be clarified below.

\section{Comparison between the two methods}

The generic case of a chiral two--dimensional theory
with central charge $c_R$ and $c_L$ for the holomorphic and 
anti--holomorphic part, respectively, is characterized by the presence 
of both diffeomorphism and trace anomaly,
\be
\nabla_\mu T^\mu {}_{\nu}=  \frac{\hbar}  {48 \pi} \frac{c_R-c_L}2\,
\epsilon_{\nu\mu}\, \partial^\mu R\label{Diffanom}
\ee
and
\be
T_\alpha^\alpha= \frac {\hbar}{48\pi}\, (c_R+c_L) \,R\label{traceanom}
\ee

Let us rewrite these equations in terms of the light--cone coordinates 
$u$ and 
$v$ introduced before. In this basis the nonvanishing metric elements 
take the form:
\be
g_{uv}=\frac 12 e^\varphi=-\epsilon_{uv},\quad\quad g^{uv}= 
2 e^{-\varphi}= \epsilon^{uv}\label{lcmetric}
\ee
and eq.\refb{Diffanom} becomes
\be
\nabla_u T_{uv} +\nabla_v T_{uu} &=&\frac{\hbar}{48 \pi}\frac{c_R-c_L}2
\epsilon_{uv} \d_u R\label{nablavv}\\
\nabla_uT_{vv}+\nabla_v T_{uv} &=&-\frac{\hbar}{48 \pi}\frac{c_R-c_L}2
\epsilon_{uv} \d_v R\label{nablauu}
\ee
On the other hand \refb{traceanom} becomes
\be
T_{uv}= \frac {\hbar}{48\pi}\, \frac{c_R+c_L}4 \,R\,e^\varphi\label{Tuv}
\ee
Replacing this with $R= -4 \d_u\d_v \varphi\,e^{-\varphi}$ in \refb{nablauu},
we get
\be
\d_v T_{uu} =   \frac {\hbar}{24\pi}\, c_R\, \d_v \ET_{uu}\label{nabla12}
\ee
where
\be
\ET_{uu} = \d_u^2\varphi -\frac 12 (\d_u\varphi)^2 \label{ET}
\ee
Integrating \refb{nabla12} we get
\be
T_{uu}(u,v) = \frac {\hbar}{24\pi}\, c_R\, \ET_{uu}(u,v) + 
T_{uu}^{(hol)}(u)\label{Tuu}
\ee
where $T_{uu}^{(hol)}$ depends only on $u$.

Similarly, integrating \refb{nablavv}, one obtains
\be
T_{vv}(u,v) = \frac {\hbar}{24\pi}\, c_L\, \ET_{vv}(u,v) + 
T_{vv}^{(a-hol)}(v)\label{Tvv}
\ee
where $\ET_{vv} = \d_v^2\varphi -\frac 12 (\d_v\varphi)^2$,
and $T_{vv}^{(a-hol)}$ depends only on $v$. The two equations
\refb{Tuu} and \refb{Tvv} are our basic result. They are equivalent 
to the two equations \refb{Diffanom} and \refb{traceanom}.

In the ``trace anomaly" method we have utilized eq.\refb{Tuu}, required
that the energy--momentum tensor be conserved and imposed the conditions
(b) of the previous section. This, in particular, amounts to requiring 
$c_R=c_L$ in the region outside the horizon. We see now that the possibility to 
integrate \refb{Diffanom} in the presence of \refb{traceanom} is 
actually insensitive to the relation between $c_L$ and $c_R$ \footnote{In other words
we can integrate the trace anomaly even if $c_R\neq c_L$. This 
is clearly only a characteristic of two dimensions}.

In the ''diff anomaly" approach we integrated \refb{Diffanom} in the near
horizon region and the conserved energy--momentum divergence away from the 
horizon. Then we imposed vanishing of energy--momentum tensor at the horizon. 
It is obvious that we used again \refb{Tuu} and \refb{Tvv} in disguise. 

It is actually possible to be more specific. We have already noticed that in the
trace anomaly method $T_{uu}(r=r_H)=0$. On the other hand we point out that 
$T_{vv}^{(a-hol)}$ is constant in $r$ and $t$, for the same reason as $T_{uu}^{(hol)}$ is,
and thus vanishes upon the request of no ingoing flux from infinity. It is also easy
to see that, if $c_R=c_L$, $\ET_{uu} = \ET_{vv}$. Therefore
$T_t^r = T_{uu}-T_{vv}$ is constant everywhere and equals the outgoing flux
\refb{flux} at infinity. Therefore the $T_t^r$ of subsection 2.2 equals 
$\hat T_t^r$ of subsection 2.1. And it is also clear that the energy--momentum
tensor vanishing at the horizon in subsection 2.1 is to be compared 
with $T_{uu}(u,v)$ of subsection 2.2.

It was important to stress the basic role of \refb{Tuu} and \refb{Tvv} because, 
as we will see, when we come to higher spin currents, it is not possible to describe
the higher flux moments by means of anomalies (either trace or diff),
but the analogues of \refb{Tuu} and \refb{Tvv} still hold and provide
the desired description.
 
It is worth at this point spending a few words about the validity of the
results obtained with the above methods in relation to the reduction
from 4 to 2 dimensions mentioned in the introduction. As pointed out there
the reduction of a free massless scalar field (interacting with the 
background metric) into an infinite set of free massless scalar fields, 
is only valid near the horizon. Away from the horizon the equations
of motion of these fields acquire a potential term. These potential terms
therefore modify eqs.(\ref{Diffanom}) and \refb{traceanom} and 
consequently \refb{Tuu} and \refb{Tvv} and account for the difference
between 2 and 4 dimensions. In the literature one can find estimates
of the effect of such modifications, see for instance 
\cite{Fabbri:2005mw}. They translate into a greybody factor that cuts
the Hawking radiation at infinity calculated above by an order of magnitude.

\section{Higher moments of the Hawking radiation and higher spin currents}

The thermal bosonic spectrum of the black hole is given by the 
Planck distribution
\be
N(\omega)= \frac {g_*}{e^{\beta\omega}-1} \label{Planck}
\ee
where $1/\beta$ is the Hawking temperature and $\omega=|k|$, the absolute value of the
momentum. 
$g_*$ is the number of physical degrees of freedom in the emitted 
radiation.
In two dimensions we can define the flux moments as follows
\be
F_n= \frac {g_*}{4\pi} \int_{-\infty}^{+\infty} dk \frac {\omega\,k^{n-2} }
{e^{\beta\omega}-1} \0
\ee 
They vanish for $n$ odd, while for $n$ even they are given by
\be
F_{2n}=\frac {1}{2\pi}\int_0^\infty d\omega \omega^{2n-1} N(\omega)=
g_*\,\frac {(-1)^{n+1}}{8\pi n} B_{2n} \k^{2n}\label{moments} 
\ee
where $B_n$ are the Bernoulli numbers 
($B_2=\frac 16,B_4=-\frac 1{30},.. $).
Therefore 
the outgoing flux (\ref{flux}) is seen to correspond to $F_2$ when
$g_*=c_R$.

The authors of \cite{IMU2} posed a very interesting question: the
outgoing flux ($=F_2$), corresponds to the integrated distribution; 
is it possible to describe the higher moments of the Hawking radiation
in the same way as we described the lowest one, by means of an 
effective field theory and, in particular, by generalizing the 
above two methods?  
They suggested that this can be done in terms of higher tensorial 
currents, which play the role of the energy--momentum tensor for
higher moments. 

In \cite{BC} an example of such currents was constructed in terms of 
an elementary
complex scalar field. If the underlying holomorphic currents satisfy
a $W_\infty$ algebra, the effective covariant currents 
were shown to describe precisely the higher moments of the Hawking
radiation. Let us briefly review the construction of \cite{BC}.

\subsection{The $W_\infty$ algebra}

Higher spin currents are expressed in terms of a single complex
bosonic field ($c=2$) and use is made of the 
$W_\infty$ algebra.  
To this end we go to the Euclidean and replace $u,v$ with
the complex coordinates $z, \bar z$. 
 
Following \cite{BK} (see also \cite{Bilal,Pope1,Pope2})
the starting point is a free complex boson having the following 
two point functions
\begin{eqnarray} \label{eqtwop}
&&	\VEV{\phi(z_1)\overline{\phi}(z_2)} = 
-\log(z_1 - z_2)\CR
&&\VEV{\phi(z_1)\phi(z_2)} = 0\CR
&&	\VEV{\overline{\phi}(z_1)\overline{\phi}(z_2)} = 0
\end{eqnarray}

The currents are defined by
\begin{equation} \label{eqdefW}
	j^{(s)}_{z\ldots z}(z) = B(s) \sum_{k=1}^{s-1} (-1)^k A^s_k 
	\NO{\partial^k_z \phi(z) \partial^{s-k}_z \overline{\phi}(z)}
\end{equation}
where
\begin{equation} \label{eqdefB}
B(s) = (-\frac i4)^{s-2}\frac{2^{s-3}s!}{(2s-3)!!}, \quad\quad 
	A^s_k = \frac{1}{s-1}\binom{s-1}{k}\binom{s-1}{s-k}
\end{equation}
They satisfy a $W_\infty$ algebra \cite{BK}. It is worth recalling
that this $W_\infty$ algebra has a unique central charge, which
corresponds to the central charge of the Virasoro subalgebra. Therefore
it has a unique basic cocycle, which is the cocycle appearing
in the Virasoro subalgebra. 

The first few currents are
\begin{eqnarray} \label{eqdefWex}
j^{(2)}_{zz} &=& -\NO{\partial_z \phi \partial_z \overline{\phi}} \CR
j^{(3)}_{zzz} &=&  \frac i2 \left( 
\NO{\partial_z \phi \partial^2_z \overline{\phi}}- 
\NO{\partial^2_z \phi \partial_z \overline{\phi}}\right) \CR
j^{(4)}_{zzzz} &=&  \frac{1}{5} \left(
\NO{\partial_z \phi \partial^3_z \overline{\phi}} - 3\! 
\NO{\partial^2_z \phi \partial^2_z \overline{\phi}} + 
\NO{\partial^3_z \phi \partial_z \overline{\phi}} \right) \CR
j^{(5)}_{zzzzz} &=&  -\frac{i}{14} \left(
\NO{\partial_z \phi \partial^4_z \overline{\phi}} - 6\!
\NO{\partial^2_z \phi \partial^3_z \overline{\phi}} + 6\!
\NO{\partial^3_z \phi \partial^2_z \overline{\phi}} - 
\NO{\partial^4_z \phi \partial_z \overline{\phi}} \right) \CR
j^{(6)}_{zzzzzz} &=&  -\frac{1}{42} \left(
\NO{\partial_z \phi \partial^5_z \overline{\phi}} - 10\!
\NO{\partial^2_z \phi \partial^4_z \overline{\phi}} + 20\!
\NO{\partial^3_z \phi \partial^3_z \overline{\phi}} - 10\!
\NO{\partial^4_z \phi \partial^2_z \overline{\phi}} + 
\NO{\partial^5_z \phi \partial_z \overline{\phi}} \right)
\end{eqnarray}
Normal ordering is defined as
\begin{equation} \label{eqdefNO}
\NO{\partial^n \phi \partial^m \overline{\phi}} = 
\lim_{z_2 \rightarrow z_1} \left\{ 
\partial_{z_1}^n \phi(z_1) \partial_{z_2}^m \overline{\phi}(z_2) - 
\partial_{z_1}^n \partial_{z_2}^m \VEV{\phi(z_1)\overline{\phi}(z_2)}
\right\}
\end{equation}
As usual in the framework of conformal field theory, the operator 
product in the RHS is understood to be radial ordered.

The current $j^{(2)}_{zz}(z) = -\NO{\partial_z \phi(z) 
\partial_z \overline{\phi}(z)}$ is proportional to the (normalized) holomorphic 
energy-momentum tensor of the model
and, upon change of coordinates $z \rightarrow w(z)$, transforms as
\begin{equation} \label{eqW2trans}
\NO{\partial_z \phi \partial_z \overline{\phi}} = 
(w')^2\NO{\partial_w \phi \partial_w \overline{\phi}}
- \frac{1}{6 } \left\{ w, z \right\} 
\end{equation}
where $ \left\{ w, z \right\} $ --- the Schwarzian derivative --- is
\begin{equation} \label{eqSchwdef}
	\left\{ w, z \right\} = \frac{w'''(z)}{w'(z)} -
 \frac{3}{2}\left( \frac{w''(z)}{w'(z)} \right)^2 
\end{equation}
 
We are interested in the transformation properties of the currents 
$j^{(s)}(u)$ when $w(z)$ is 
\begin{equation}
	\label{eqdefw}
	w(z) = -e^{-\kappa z}
\end{equation}
In \cite{BC} we obtained 
\begin{equation}\label{eq2}
j^{(s)}_{z\ldots z}(z) \rightarrow  
\left( \frac 1{\kappa w}\right)^s  \left(j^{(s)}_{z\ldots z}
+\VEV{ X}_s\right) 
\end{equation}
where 
\begin{eqnarray}\label{eqXw}
	 	\VEV{X_s}= (-)^{s-1} (-i)^{s-2}
	{\kappa^s} \frac{B_s}{s}
\end{eqnarray}
Eq.\refb{eq2} has to be compared with eq.\refb{TholU}.
This is a higher order Schwarzian derivative evaluated at 
$w(z) = -e^{-\kappa z}$. It plays a role analogous to the RHS of 
\refb{flux}. Below we will compare it with the
radiation moments in the RHS of \refb{moments}.

\subsection{Higher spin covariant currents}

Let us now return to the light--cone notation. We identify 
$j^{(2)}_{uu}(u)$ up to a constant with
the holomorphic energy momentum tensor 
\be
 j_{uu}^{(2)}(u)=-2\pi\, T^{(hol)}_{uu}\label{j2T}
\ee
Similarly we identify $j^{(s)}_{u\ldots u}$, with $s$ lower 
indices, with an $s$--th order holomorphic tensor. They can be 
naturally thought of as the only
non--vanishing components of a two--dimensional completely 
symmetric current. In analogy with the
energy--momentum tensor, we expect that there exist
a conformally covariant version $J^{(s)}_{u\ldots u}$ of 
$j^{(s)}_{u\ldots u}$. 
The latter must be the intrinsic 
component of a two--dimensional completely symmetric traceless current
$ J^{(s)}_{\mu_1\ldots\mu_s}$, whose only other classically non--vanishing
component is $J^{(s)}_{v\ldots v}$. We identify them with the 
currents \refb{Js}

The previous holomorphic currents refer to a background with trivial (Euclidean) 
metric. In order to find a covariant expression of them we have
to be able to incorporate the information of a non--trivial metric.
This was done in \cite{BC} following \cite{IMU2}.  
According to the recipe explained there, the covariant counterpart
of  $j^{(s)}_{u\ldots u}$ should be constructed using currents
\begin{equation}
J^{(n,m)}_{u\ldots u} = e^{(n+m)\varphi(u)}\lim_{\epsilon \rightarrow 0} 
\left\{ e^{-n\varphi(u_1)-m\varphi(u_2)} 
\nabla^n_{u_1} \phi \nabla^m_{u_2} \overline{\phi} - 
\frac{c_{n,m} \hbar}   {\epsilon^{n+m}}\right\}\label{Jnm}
\end{equation}
where $c_{m,n} = (-)^{m} ( n+m-1 )!$ are numerical constants determined in such a way
that all singularities are canceled in the final expression for $J^{(n,m)}_{u\ldots u}$.
Therefore
\refb{Jnm} defines the normal ordered current
\begin{equation}
	J^{(n,m)}_{u\ldots u} = \NO{\nabla^n_u \phi \nabla^m_u 
\overline{\phi}}\label{Jnmno}
\end{equation}
 
After some algebra one gets

\begin{eqnarray}\label{Jsjs}
	J^{(2)}_{uu} &=& j^{(2)}_{uu}  - \frac{ \hbar }{6}  \ET  \CR
	J^{(3)}_{uuu} &=&  j^{(3)}_{uuu}  \CR
	J^{(4)}_{uuuu} &=&  j^{(4)}_{uuuu}  + \frac{ \hbar}{30}   \ET^2 +
\frac{2}{5} \ET J^{(2)}_{uu}   \CR 
	J^{(5)}_{uuuuu} &=&  j^{(5)}_{uuuuu} + \frac{10}{7}   \ET  J^{(3)}_{uuu}  
\end{eqnarray}
and
\begin{eqnarray}\label{J6j6}
	J^{(6)}_{uuuuuu} &=&  \left(
-\frac{2 \hbar }{63  }\ET^3
+\frac{5 \hbar }{504  }\left(\partial _u\ET\right)^2
-\frac{\hbar }{126  }\ET\partial _u^2\ET \right. 
\CR
&&
-\frac{2}{3}\ET^2J^{(2)}_{uu}
-\frac{1}{21}\ET\nabla _u^2J^{(2)}_{uu}
-\frac{1}{21}\left(\partial _u^2\ET\right)J^{(2)}_{uu}
+\frac{5}{42}\left(\partial _u\ET\right)\nabla _uJ^{(2)}_{uu}
\CR
&&
\left.
-\frac{5}{21}\Gamma \ET\nabla _uJ^{(2)}_{uu}
-\frac{5}{21}\Gamma ^2\ET J^{(2)}_{uu}
+\frac{5}{21}\Gamma \left(\partial _u\ET\right)J^{(2)}_{uu}\right) 
-\frac{5}{24}\ET J^{(4)}_{uuuu}
+j^{(6)}_{uuuuuu}
\end{eqnarray}
where
\begin{equation}
\ET = \partial^2_u \varphi -  \frac{1}{2} 
\left( \partial_u \varphi \right)^2  
\end{equation}
These equations are the analogs of \refb{Tuu}.

The covariant divergences of these currents are
\begin{eqnarray}
 g^{uv}\nabla _v	J^{(2)}_{uu} &=&  
\frac{ \hbar }{12} \left(\nabla _uR\right) 
\label{djv2}\\
g^{uv}\nabla _v J^{(3)}_{uuu} &=& 0
\label{djv3}\\
 g^{uv}\nabla _v	J^{(4)}_{uuuu} +  \frac{ 1}{5}  q^2 
\left(\nabla _uR\right) J^{(2)}_{uu}&=& 0 \label{djv4} 
\\
g^{uv}\nabla _v  J^{(5)}_{uuuuu}  +  
\frac{5}{7}  \left(\nabla _uR\right) J^{(3)}_{uuu}&=& 0
\label{djv5} \end{eqnarray}
and, for  $s=6$
\begin{eqnarray}\nonumber
g^{uv}\nabla _v  J^{(6)}_{uuuuuu}  &+&
\left(\frac{5}{84}\left(\nabla _u^2R\right)\nabla _uJ^{(2)}_{uu}
-\frac{1}{42}\left(\nabla _uR\right)\nabla _u^2J^{(2)}_{uu}
-\frac{1}{42}\left(\nabla _u^3R\right)J^{(2)}_{uu}\right) 
\label{djv6}\\
&+& \frac{5}{3}\left(\nabla _uR\right)J^{(4)}_{uuuu} =0
\end{eqnarray}
Eq.\refb{djv2} is to be compared with \refb{nabla12} while the remaining
ones are the relevant higher spin analogs. 

The above equations mean that all the higher spin equations are covariantly
conserved. In the RHS of \refb{djv3}--\refb{djv6}, unlike \refb{djv2},
there does not appear any terms proportional to $\hbar$. Any such term 
must be interpreted as the consequence of a trace anomaly (and possibly
a diff anomaly) as has been argued by \cite{IMU4}. In other words if 
there is a term proportional to $\hbar$ in $g^{uv} 
\nabla_v J_{uu\ldots u}$ this must be understood as related to the
second term in the covariant divergence $\nabla^\mu  J_{\mu u\ldots u} = g^{uv} 
\nabla_v J_{uu\ldots u} + g^{uv} \nabla_u J_{vu\ldots u}$. Such a term 
tells us that $J_{vu\ldots u}$, which classically vanishes, takes on a nonzero
value at one loop, revealing the existence of a trace anomaly. 
This is precisely what happens for the covariant second order current
(energy--momentum tensor) $J^{(2)}_{\mu\nu}$ \refb{djv2}: the trace is  
$\Tr( J^{(2)}{} ) = 2 g^{vu}J^{(2)}{}_{vu} $. 
Thus, \refb{djv2} reproduces the well 
known trace anomaly $ \Tr( J^{(2)} ) = -\frac{c \hbar}{12} R$, where 
in our case $c_R=2$ \footnote{We relate $j^{(2)}_{uu}$ to the energy 
momentum 
tensor via the factor of $2\pi$ and the minus sign. This is because  
in the Euclidean we want to conform to the conventions and results of
\cite{BK}, where properly normalized currents satisfy a $W_\infty$ algebra.
This holds for higher order currents too: 
for physical applications their $W_\infty$ representatives must 
all be divided by $-2\pi$.}.

However for the other equations, we see that the terms that carry explicit 
factors of $\hbar$ cancel out
in eqs.~\refb{djv3}-\refb{djv6}. This implies the absence of $\hbar$ 
terms in the trace, and consequently the absence of any trace anomaly as well as
of any diffeomorphism anomaly.

In \cite{BC} it was shown that, as far as trace anomalies are concerned,
this result is to be expected, since
via a cohomological analysis it can be seen that no true trace anomaly 
can exist in higher spin currents.
 
Of course we could repeat the same construction for antiholomorphic
currents and find the corresponding covariant ones. We would find
perfectly symmetric results with respect to the ones above.

\subsection{Higher moments of the Hawking radiation}

Now let us apply to the just introduced higher spin currents 
an argument similar to the one in section 2 
for the energy--momentum tensor, using the previous results from 
the $W_\infty$ algebra. Introducing the Kruskal coordinate 
$U=-e^{-\kappa u}$ and 
requiring regularity at the horizon we find that, at the horizon, the value
of $j^{(s)}_{u\ldots u}$ is given by $\VEV{X_s}$ in eq.\refb{eqXw}. Next
$j^{(s)}_{u\ldots u}(u)$ is constant in $t$ and $r$ (the same is of course true
for $j^{(s)}_{v\ldots v}$). Therefore, if we identify 
$j^{(s)}_{u\ldots u}(u)$ with $j^{(s)}_{z\ldots z}(z)$ via Wick rotation,
$\VEV{X_s}$ corresponds to its value at $r=\infty$. Since $j^{(s)}_{u\ldots u}(u)$
and  $J^{(s)}_{u\ldots u}(u)$ asymptotically coincide, the asymptotic
flux of these currents is
\be
-\frac 1{2\pi} \VEV{J^{(s)^r}{}_{t\ldots t}}= 
-\frac 1{2\pi} \VEV{J^{(s)}_{u\ldots u}}+
\frac 1{2\pi} \VEV{J^{(s)}_{v\ldots v}}= -\frac 1{2\pi} \VEV{X_s}= 
\frac {i^{s-2}}{2\pi s} \kappa^s B_s \label{sflux}
\ee
For the global $-2\pi$ factor, see the previous footnote. 

The RHS vanishes
for odd $s$ (except $s=1$ which is not excited in our case) and coincides 
with the thermal flux moments \refb{moments} for even $s$.

\subsection{A qualitative motivation for higher spin currents}

We would like to spend a few words concerning the origin of  
higher spin currents, even though what follows is very qualitative 
and is in fact not needed in the economy of the paper. 

Let us suppose we know the energy 
momentum tensor of a fundamental theory which faithfully reproduces
the full spectrum of the Hawking radiation and expand it around
our background metric. To guess what may occur think of a quantum 
energy--momentum 
tensor represented in the Sugawara form in a flat background:
$T_{\mu\nu}= (:J_\mu J_\nu:- {\rm trace})$, where, for instance,  
$J_\mu=\d_\mu \phi$ in the simplest case. We can view it as an
expression point-split by a small but finite amount $2y$
\be
T_{\mu\nu}(x)= \lim_{y \to 0} :\d_\mu \phi(x-y) \d_\nu
\phi(x+y)- {\rm trace}:\label{pointsplit}
\ee
The finite point splitting is meant to account for a nonlocal 
interaction that synthesizes the interactions of the underlying model
(see the related considerations in \cite{IMU2}). 
Let us expand in Taylor series
\be
&&:\d_\mu \phi(x-y) \d_\nu \phi(x+y):\0\\
&&\quad = \sum_{i=0} \sum_{j=0}\frac {(-1)^i}{i!j!}
:y^{\mu_1}\ldots y^{\mu_i}\d_{\mu}\d_{\mu_1}\dots \d_{\mu_i}\phi(x)
y^{\nu_1}\ldots y^{\nu_j}\d_{\nu}\d_{\nu_1}\dots \d_{\nu_j}\phi(x):
\label{pointsplit1}
\ee
This expansion is appropriate for a two-dimensional flat space--time,
but we will need to consider point splitting in a curved space--time.
Therefore in \refb{pointsplit1} the derivative will be
replaced by covariant derivative and the products 
$y^{\mu_1}\ldots y^{\mu_i}y^{\nu_1}\ldots y^{\nu_j}$
by complicated expressions of the background. We represent all this by 
effective background tensor fields $B^{(s)}_{\mu_1\ldots\mu_s}$.
When inserted back in \refb{pointsplit}, the quantum expression will 
give rise to an expansion of the energy--momentum tensor in terms of 
higher spin currents coupled to such fields.

In a previous subsection we have 
constructed higher spin currents from a $W_\infty$ algebra using
a chiral coordinate $z$, which we understand as the local holomorphic 
coordinate over a Riemann surface $\Sigma$. A $W_\infty$ algebra is 
generated on a local patch not only by diffeomorphisms, but by more 
general
coordinate transformations, the symplectomorphisms, which involve also 
the cotangent bundle of $\Sigma$, see \cite{hull} and, 
for an explicit construction, \cite{BZ}. In particular in \cite{BZ} it 
is shown that from general transformations of the type
\be
\delta C^{(r)}(z,\bar z)= \sum_{s=1}^r s \,C^{(s)} (z,\bar z)\,\d_z 
C^{(r-s-2)}(z,\bar z)\label{Wtransf} 
\ee
where $C^{(r)}$ are `ghost' tensors of order $r$,
the following algebra follows for an infinite set of generators 
$T^{(r)}(z,\bar z)$,
\be
\left[T^{(r)}(z,\bar z), T^{(s)}(z',\bar z')\right]=
(r-1) \d^{z'}\delta (z'-z) T^{r+s-2}(z,\bar z)-
(s-1) \d_z \delta(z- z')T^{r+s-2}(z',\bar z')
\0
\ee
This is the classical version of the $W_\infty$ algebra (for a 
quantum version, see for instance \cite{BK}). It is possible to 
recognize in \refb{Wtransf} the transformations \refb{deltaxixi} 
and \refb{deltaxitau} below. 

Now the above expression \refb{pointsplit} exhibits a dependence both on
$x^\mu$ and on $y^\mu\equiv dx^\mu$. We can think of $y^\mu$ as 
local coordinates on the cotangent bundle of $\Sigma$ and their 
transformations can be conceived of as $W_\infty$ transformations. 
Therefore, even though the details fully depend on the fundamental 
theory and remain implicit, the appearance of higher spin currents
and their $W_\infty$ algebra structure is not so surprising.

Each of these higher spin currents carries to infinity its own piece of 
information about the Hawking radiation. Just in the same way as in the action 
the metric is a source for the energy--momentum tensor, these new 
(covariant) currents
will have in the effective action suitable sources, with  the appropriate
indices and symmetries. In \cite{BC} they were represented by
asymptotically trivial
background fields $B^{(s)}_{\mu_1\ldots\mu_s}$ (in \cite{hull} they were 
called `cometric functions'). So we have
\be
 J^{(s)}_{\mu_1\ldots\mu_s}= \frac 1{\sqrt{g}} \frac {\delta}{\delta 
B^{(s)\mu_1\ldots\mu_s}} S\label{Js}
\ee
In particular $B^{(2)}_{\mu\nu}= g_{\mu\nu}/2$. We assume that
all $J^{(s)}_{\mu_1\ldots\mu_s}$ are maximally symmetric and classically
traceless.

\section{Diffeomorphism anomalies for higher spin currents}

In subsection 4.2 we saw that it is consistent to require
that higher spin currents are covariantly conserved. This leads 
for such higher tensor currents to the absence both of trace and 
diffeomeorphisms anomalies. The trace and covariant divergence of
the currents were determined with a particular construction based on
currents made out of a bosonic scalar field. 
Therefore it is important to find an independent confirmation of such 
results. 

As for the trace anomalies it was shown in \cite{BC} that 
this is no accident: the trace of the fourth 
order current does not admit true anomalies (there may appear anomalous 
terms, but they correspond to trivial cocycles and can be canceled by 
suitable counterterms
in the effective action). This result is seemingly valid for all the
higher spin currents, because a thumb rule suggests that true anomalies
appear only when the cocycle engineering dimension (in our case the
total number of derivatives) is related in a precise way to the
space--time dimension.

As for the diffeomorphism anomalies, on the basis of the previous construction
there is no evidence of them either. But in \cite{IMU2} some
diff anomalies appeared in the covariant divergence of higher spin 
(bi--spinorial) currents. It is therefore important to verify that 
this is not in contrast with our results above. This means that we have 
to prove that such anomalies are trivial. Eqs.\refb{djv2} through
\refb{djv6}, are covariant conservation equations (as it is apparent
in eq.\refb{djv2}). Therefore, if anomalies ever appear in such
conservation equations, they appear in covariant form. Existence or
non--existence of covariant anomalies is not easy to analyze in general,
while general results can be obtained for consistent anomalies. Since
absence of consistent anomalies implies absence of the corresponding covariant ones,
we will try to show that, for the conservation laws we are interested in,
there are no consistent anomalies (except the well--known one 
corresponding to \refb{djv2}). It should be remarked that this
problem is interesting in itself, even independently of the 
application considered in this paper, and, to our best knowledge,
has not been studied so far.

In the sequel we will give for the fourth order current a proof 
of absence of diff anomalies analogous to the one that was 
presented in \cite{BC} for trace anomalies and, under reasonable 
assumptions, we will extend the proof to currents of any order. 
This will lend support to our previous claims, beyond the 
explicit construction of the previous section.
 
\subsection{The consistency method for diff anomalies}

The conservation of the energy--momentum tensor corresponds, as is well--known,
to the symmetry of the theory under the diffeomorphism transformations:
\be
\delta_\xi g_{\mu\nu}= \nabla_\mu \xi_\nu + \nabla_\nu \xi_\mu\label{deltaximu}
\ee
where $\xi_\mu = g_{\mu\nu} \xi^\nu$, and $\xi^\mu$ represent
infinitesimal general coordinate transformations $x^\mu \to x^\mu + \xi^\mu$.
The background fields transform in a covariant way under these transformation
\be
\delta_\xi B^{(s)}_{\mu_1\ldots\mu_s}= \xi^\l \d_\l B^{(s)}_{\mu_1\ldots\mu_s}
+ \d_{\mu_1} \xi^\l B^{(s)}_{\l\ldots\mu_s}+\ldots + 
 \d_{\mu_s} \xi^\l B^{(s)}_{\mu_1\ldots\l}\label{deltaBsxi}
\ee 

Similarly the conservations of higher spin currents correspond to the symmetry
under higher tensorial transformation. In particular the conservation of 
$J^{(4)}$ is due to invariance under
\be
\delta_\tau B^{(4)}_{\mu_1\mu_2\mu_3\mu_4}= 
\nabla_{\mu_1} \tau_{\mu_2\mu_3\mu_4} + { cycl.}\label{deltaBtau}
\ee
where $\tau$ is a completely symmetric traceless tensor and ${ cycl}$ 
denotes cyclic permutations of the indices. The reason for tracelessness
will be given later.

To find the (consistent) anomalies of the energy--momentum tensor and 
higher spin currents with respect to the symmetry induced by the 
above transformations, we will analyze the solutions of the relevant 
Wess--Zumino consistency conditions. An equivalent (and simpler) way is 
to transform the problem into a cohomological one. The trick 
is well--known. We promote the transformation parameters to anticommuting 
ghost fields and endow them with a suitable transformation law. 
This gives rise to a nilpotent operator acting on the local functionals
of the fields and their derivatives. Local functionals (cochains) and
nilpotent operator (coboundary) define a differential complex.
Anomalies correspond to non--trivial cocycles.

For $\xi$ this leads to
\be
\delta_\xi\xi^\mu= \xi^\l \d_\l \xi^\mu \label{deltaxixi}
\ee
beside
\be
\delta_\xi\tau_{\mu\nu\rho}=\xi^\l \d_\l \tau_{\mu\nu\rho}+ \d_\mu\xi^\l \tau_{\l\nu\rho}+
\d_\nu\xi^\l \tau_{\mu\l\rho}+\d_\rho\xi^\l \tau_{\mu\nu\l} \label{deltaxitau}
\ee
It is then easy to show that $\delta_\xi^2=0$. 

In a similar way, beside $\delta_\tau g_{\mu\nu}=0$, we set
\be
\delta_\tau \tau_{\mu\nu\l}=0\label{deltatautau}
\ee
that is, we assume that $\tau$ is an Abelian parameter. This is not 
obvious a priori and requires a specific justification. We do it 
in Appendix A. Now it is elementary to prove that, as a consequence of 
the anticommutativity of $\tau$ we have $\delta_\tau^2=0$. More generally,
since $\tau$ is assumed to anticommute with $\xi$, we have
\be
\delta_\xi^2=0,\quad\quad \delta_\tau^2=0,\quad\quad 
\delta_\xi\delta_\tau+\delta_\tau \delta_\xi=0\label{delta2}
\ee
In the following we will denote by $\delta_\tau,\delta_\xi$ also the 
corresponding functional operators. It follows from \refb{delta2} 
that the operator $\delta_{tr} = \delta_\xi+\delta_\tau$ is nilpotent.
It is clear that $\delta_{tr}$ is not the total functional operator
of our system, but rather a truncated one, since we are disregarding
higher tensorial gauge transformations\footnote{To be precise we are  
concentrating on eqs.(\ref{djv2},\ref{djv4}) and disregarding \refb{djv6}.
The conservation laws \refb{djv3} and \refb{djv5} do not admit anomalies 
in the present context.}. Such a truncation is justified by the fact that
our differential system is graded. This can be seen as follows.

Let us recall first the canonical dimensions of the various fields 
involved. $g_{\mu\nu}$ has dimensions 0; $\xi$ has dimension (in mass) -1,
while $B^{(4)}$ and $\tau$ have dimensions -2 and -3, respectively.
Now let us consider the nilpotent total differential operator 
$\delta_{tot}= \delta_\xi+\delta_\tau+\ldots$. Then 
(integrated) anomalies are defined by
\be
\delta_{tot}\, \Gamma^{(1)} = \hbar \Delta,\quad\quad \delta_{tot} \Delta=0
\label{defanom}
\ee
where $\Gamma^{(1)}$ is the one--loop quantum action. $\Delta$, which is
the integral of a local functional in the fields and their derivatives,
splits naturally into $\Delta_\xi+\Delta_\tau+\ldots$. In turn each 
addend splits into a sum of terms according to the degree of their 
integrand. The degree is defined by the number of derivative of the
integrand minus 1. Therefore we have for instance
\be
\Delta_\xi=\Delta_\xi^{(2)} +  \Delta_\xi^{(4)}+\Delta_\xi^{(6)}+\ldots,
\quad\quad \Delta_\tau=\Delta_\tau^{(4)}+\Delta_\tau^{(6)}+\ldots\0
\ee
As a consequence $\delta_{tot} \Delta=0$  splits into
\be
&&\delta_\xi \Delta_\xi^{(2)}=0 \label{deltaxi2}\\
&&\delta_\xi \Delta_\xi^{(4)}=0 \label{deltaxi4}\\
&&\delta_\xi \Delta_\xi^{(6)}=0, \quad\ldots \label{deltaxi6}
\ee
and
\be
&&\delta_\tau \Delta_\tau^{(4)}=0\label{deltatau4}\\
&&\delta_\tau \Delta_\tau^{(6)}=0,\quad \ldots\label{deltatau6}
\ee
with the cross conditions\footnote{The action of $\delta_\tau$ on 
$\Delta_\xi^{(2)}$ is trivial.}
\be
&&\delta_\tau\Delta_\xi^{(4)}+ \delta_\xi \Delta_\tau^{(4)}=0\label{cross4}\\
&&\delta_\tau\Delta_\xi^{(6)}+ \delta_\xi \Delta_\tau^{(6)}=0,\quad\ldots\label{cross6}
\ee
Therefore, fortunately, our complex splits 
into subcomplexes and, for example it makes sense to truncate it
at level 4, i.e. to eqs.(\ref{deltaxi2},\ref{deltaxi4},\ref{deltatau4})
and \refb{cross4}, since these conditions are not affected by the
higher order equations in the complex. 

\subsection{The search for $\delta_\tau$ anomalies}

Let us explain the strategy to prove the absence of anomalies for fourth 
order currents. The first step is to solve 
eqs.(\ref{deltaxi2},\ref{deltaxi4}) in general. We will show that, while
\refb{deltaxi2} admits a nontrivial solution (the 2d diff anomaly),  
\refb{deltaxi4} does not admit any nontrivial solution. This will be
done in Appendix B: the proof is based on an argument used for 
4d anomalies in \cite{BPT} and adapted to the present context. What we 
prove precisely is that any solution to eq.\refb{deltaxi4} is trivial,
that is there exist a local functional $C^{(4)}$ of the background fields 
such that if $\Delta_\xi^{(4)}$ is a solution to \refb{deltaxi4}, then
$\Delta_\xi^{(4)}= \delta_\xi \, C^{(4)}$. Therefore we can rewrite
\refb{cross4} as 
\be
\delta_\xi (\Delta_\tau^{(4)} -\delta_\tau C^{(4)})=0\label{deltatau4'}
\ee
This amounts to saying that any cocycle of $\delta_\tau$ 
(i.e. any solution
to \refb{deltatau4}) can be written in a diff--covariant form.
This is a piece of very useful information because it strongly limits the forms
of the cochains we have to analyze in order to find the solutions to
\refb{deltatau4}.

What remains for us to do is very simple. Let us start with an example.
We write a first set of chains
\be
\Delta_\tau = \int d^2x \,\sqrt{-g} \sum_{i=1}^3 a_i\, I^\tau_i 
\label{Deltatau}
\ee
where
\be
I^\tau_1= \tau^{\mu\nu\l}\, \nabla_\mu \nabla_\nu \nabla_\l R,\quad\quad
I^\tau_2= \tau^{\mu\l}{}_\l\, \square\nabla_\mu R, \quad\quad
 I^\tau_3= \tau^{\mu\l}{}_\l\, \nabla_\mu R^2\0
\ee
where we have ignored tracelessness of $\tau$.
All these cochains are, trivially, cocycles of $\delta_\tau$ and they are the only
ones one can construct of this type\footnote{In $2$ dimensions we have
\begin{eqnarray}
	R_{\mu \nu \lambda \rho} &=&  \frac{1}{2} R 
\left( g_{\mu \lambda}\, g_{\nu \rho} 
- g_{\mu \rho}\, g_{\nu \lambda} \right)  \CR
	R_{\mu \nu}  &=&  \frac{1}{2} g_{\mu \nu}\, R 
\end{eqnarray}
}.

Next we have to find out whether these cocycles are trivial or not.
The only possible counterterms are also 3.
\be
C= \int d^2x \, \sqrt{-g} \sum_{j=1}^3 c_j J_j\label{counterC}
\ee
where
\be
J_1= B^{\mu\nu\l}{}_\l\, \nabla_\mu\nabla_\nu R,\quad\quad
J_2= B^{\mu\l}{}_{\mu\l}\, \nabla^\nu\nabla_\nu R,\quad\quad 
J_3= B^{\mu\l}{}_{\mu\l}\, R^2\0
\ee

Applying $\delta_\tau$ to \refb{counterC} we get
\be
\delta_\tau C =  \int d^2x \, \sqrt{-g} \sum_{i,j=1}^3 c_i M_{ij} I^\tau_j
\label{deltatauC}\ee 
where $M_{ij}$ is the matrix
\be
M_{ij} =- \left(\begin{matrix} 2 & 2 & 0\\
                               0& 4&-1\\
                               0&0& 4 \end{matrix}\right)
\ee
Since the determinant of this matrix is nonvanishing we can always find
$c_i$ such that \refb{deltatauC} reproduce \refb{Deltatau} for any choice of
the parameters $a_i$. Therefore all the cocycles \refb{Deltatau}
are trivial. 

This is not enough since the cocycles \refb{Deltatau}
are not of the most general form. We expect 
a true diff anomaly to contain the $\epsilon_{\mu\nu}$ tensor (see section 3). 
There are three cochains of such a form
\be
\Delta_\tau = \int d^2x \,\sqrt{-g} \sum_{i=1}^3 b_i\, K^\tau_i 
\label{Deltataueps}
\ee
where
\be
K^\tau_1= \tau^{\mu\nu\l}\, 
\epsilon_{\mu\alpha} \nabla^\alpha \nabla_\nu \nabla_\l R,
\quad\quad
K^\tau_2= \tau^{\mu\l}{}_\l\, 
\epsilon_{\mu\alpha}\square\nabla^\alpha R, \quad\quad
K^\tau_3= \tau^{\mu\l}{}_\l\,\epsilon_{\mu\alpha} \nabla^\alpha  R^2
\label{K123}
\ee
They are, trivially, cocycles.

On the other hand now there is only one possible counterterm
\be
C= \int d^2x \, \sqrt{-g} B^{\mu\nu\l}{}_\l\, \epsilon_{\nu\alpha} 
\nabla^\alpha\nabla_\mu R \label{counterCeps}
\ee
It it easy to see that
\be
\delta_\tau C= -\int d^2x \, \sqrt{-g}\left( 2 \tau^{\mu\l\rho}
 \epsilon_{\rho\alpha}\, \nabla^\alpha\nabla_\l \nabla_\mu R +
\tau^{\mu \l}{}_{\l} \epsilon_{\mu\alpha}\, \square \nabla^\alpha R+
2 \tau^{\mu \l}{}_{\l}\epsilon_{\mu\alpha}\,R \nabla^\alpha R\right)
\label{deltatauCeps}
\ee 
Therefore this counterterm is not enough to cancel the three previous 
independent cocycles. Here come tracelessness of $\tau$. This property
is necessary because it is easy to realize that the last two terms
in \refb{K123}, which are proportional to $ \tau^{\mu \l}{}_{\l}$, would 
appear in conservation laws in which also the components $J^{(4)}_{uuvv}$
are `excited'. This would bring us outside our system. To avoid this we
have to impose that $\tau$ is traceless. This being so, only $K_1^\tau$
survives among the cocycles, and only the first term survives in the
RHS of \refb{deltatauCeps}. The latter precisely cancels the only possible 
nontrivial cocycle.

To conclude, there are no non--trivial consistent anomalies in the divergence
of the fourth order current.

It is not hard to extend the above argument to sixth and higher order currents,
provided we assume that all the chains can be written in a covariant form. 
This corresponds to assuming that there are no non--trivial solutions to 
eq.\refb{deltaxi6} and the analogous higher equations. Proving this result requires 
a refinement of the techniques used in Appendix B, and we will not do it here.
However it is very reasonable to assume it.

Let us prove the following claim: all solutions to the equation
$$\delta_{\omega} \Delta^{(2n)}_{\omega} = 0$$
where $\omega^{\mu_{1} \dots \mu_{2n-1}}$ is a totally symmetric, traceless 
ghost parameter (the generalization of $\tau_{\mu\nu\l}$), are trivial, 
i.~e.~there exists a local functional $C^{(2n)}$ of the 
background fields, such that $$\Delta^{(2n)}_{\omega} = \delta_{\omega} C^{(2n)}$$.

We will show this under the assumption that all chains $\delta_{\omega}$ acts upon 
can be written in a diff-covariant form. Therefore we start by writing the most general
cocycles as  
\be
\Delta_{\omega} = \int d^{2}x \sqrt{-g} \left( a I^{\omega} + b K^{\omega} \right)
\label{Deltaomega}
\ee
where $a$ and $b$ are constants and $I^{\omega}$ and $K^{\omega}$ are 
the {\it only} possible terms (see Appendix B) we can construct in $D=2$, 
taking into account the tracelessness of $\omega$. Their explicit form is:
\be
I_{\omega} = \omega^{\mu_{1} \dots \mu_{2n-1}} 
\nabla_{\mu_{1}} \dots \nabla_{\mu_{2n-1}} R\label{Iomega}
\ee
and
\be
K_{\omega} = \omega^{\mu_{1} \dots \mu_{2n-1}} \, \epsilon_{\mu_{1} \alpha} \, 
\nabla^{\alpha} \nabla_{\mu_{2}} \dots \nabla_{\mu_{2n-1}} R\label{Komega}
\ee
\noindent
Now we claim that the corresponding counterterm is the following:
\be
C = -\frac{1}{2} \int d^{2}x \sqrt{-g} \left( a J^{\omega} + b L^{\omega} \right)
\label{contC}
\ee
where
\be
J_{\omega} = {B}^{\mu_{1}\dots\mu_{2n-2}\sigma}{}_\sigma \, 
\nabla_{\mu_{1}} \dots \nabla_{\mu_{2n-2}} R\label{Jomega}
\ee
and
\be
L_{\omega} = {B}^{\mu_{1}\dots\mu_{2n-2}\sigma}{}_\sigma \, \epsilon_{\mu_{1} 
\alpha} \nabla^{\alpha} \nabla_{\mu_{2}} \dots \nabla_{\mu_{2n-2}} R\label{Lomega}
\ee
and $B$ is the corresponding background field. Using the formulas
\be
&&\delta_{\omega} g_{\mu \nu} = 0\0\\
&&\delta_{\omega} B^{\mu_{1} \dots \mu_{2n}} = 
\nabla^{\mu_{1}} \omega^{\mu_{2} \dots \mu_{2n}} + \textrm{cycl.}\0
\ee
and again the fact that $\omega$ is traceless, we get, after integration by parts,
\be
\delta_{\omega} C &=& \int d^{2}x \sqrt{-g} \left( a \, \omega^{\mu_{1} \dots 
\mu_{2n-2} \sigma} \nabla_{\sigma} \nabla_{\mu_{1}} \dots \nabla_{\mu_{2n-2}} R \ 
+ \right.\0\\
&+&\left.  b \, \omega^{\mu_{1} \dots \mu_{2n-2} \sigma} 
\epsilon_{\mu_{1} \alpha} \nabla_{\sigma} \nabla^{\alpha} \nabla_{\mu_{2}} 
\dots \nabla_{\mu_{2n-2}} R \right)\0
\ee
The second term under the integral has to be rearranged by reversing the order of 
the first two covariant derivatives ($\nabla_{\sigma} \nabla^{\alpha}$). 
Using the formulas in the previous footnote, we have      
\be
&&\omega^{\mu_{1} \dots \mu_{2n-2} \sigma} \epsilon_{\mu_{1} \alpha} \nabla_{\sigma} 
\nabla^{\alpha} \nabla_{\mu_{2}} \dots \nabla_{\mu_{2n-2}} R = 
\omega^{\mu_{1} \dots \mu_{2n-2} \sigma} \epsilon_{\mu_{1} \alpha} \nabla^{\alpha} 
\nabla_{\sigma} \nabla_{\mu_{2}} \dots \nabla_{\mu_{2n-2}} R \, \0\\
&&\quad\quad+ \, \omega^{\mu_{1} \dots \mu_{2n-2} \sigma} \, \epsilon_{\mu_{1} \alpha} \,
{R}_\sigma{}^\alpha{}_{\mu_{2}}{}^\lambda \nabla_{\lambda} \nabla_{\mu_{3}} 
\dots \nabla_{\mu_{2n-2}} R \, + \dots\0
\ee
So, a typical additional term has a form
\be
\frac{R}{2} \, \omega^{\mu_{1} \dots \mu_{2n-2} \sigma} \, \epsilon_{\mu_{1} \alpha} \, 
(g_{\sigma \mu_{i}} \, g^{\alpha \lambda} - 
g_{\sigma}^{\lambda} \, g^{\alpha}_{\mu_{i}}) \nabla_{\mu_{2}} \dots \nabla_{\mu_{i-1}} 
\nabla_{\lambda} \nabla_{\mu_{i+1}} \dots \nabla_{\mu_{2n-2}} R\0
\ee
The first part of this term vanishes because of the tracelessness of $\omega$ 
and the second because it leads to contraction of antisymmetric $\epsilon$ tensor 
and symmetric indices in $\omega$. Therefore, we have proven that
\be
\delta_{\omega} C = \int d^{2}x \sqrt{-g} \left( a I^{\omega} + b K^{\omega} \right)
\label{trivialomega}
\ee
This means that, allowing for the above assumption,  
there are 
no non--trivial anomalies in any higher spin currents. 
Therefore a properly chosen 
regularization should not produce any covariant anomaly either. 
This is reflected in our eqs.\refb{djv4} and \refb{djv6}, which express the covariant
conservation of the fourth and sixth order currents. The additional terms in the LHS
(which are not present in the consistent version of the conservation law)
are needed in order to guarantee covariance of the divergence in 
the presence of the non--trivial gravitational background (see Appendix A).

\section{Conclusions}
 
In this paper we have shown that the two methods of calculating the integrated
flux of Hawking radiation on a static symmetric black hole, the method that makes 
use of the trace anomaly and the one based
on the  diffeomorphism anomaly, are strictly related. The two methods actually 
boil down to the same basic elements. We have also pointed out the basic role
of the integrated conservation equations \refb{Tuu} and \refb{Tvv}.

In order to describe the higher moments of the Hawking radiation spectrum, we have
introduced higher spin currents. They have been constructed starting from
a $W_\infty$ algebra on the complex plane and subsequently lifted to the curved space--time
corresponding to the black hole background metric. They were shown in \cite{BC}
to describe the higher moments of the black hole emission. We passed then to analyze
the presence of anomalies in the traces and covariant divergences of these
higher tensorial currents. The above mentioned explicit construction reveals none. 
Therefore we went on to analyze the possible existence of higher order trace and diff
anomalies, relying on consistency methods (Wess--Zumino consistency conditions).
In \cite{BC} it was shown that no trace anomaly exists for the fourth order current.
In this paper we have analyzed the most challenging problem of diff anomalies.
The result is still negative: no non--trivial anomalies exist.

The extension of the anomaly analysis to still higher orders is very challenging, but we
believe that we have gathered enough evidence that higher spin currents cannot
have anomalies, only the energy--momentum tensor can. This corresponds to a prejudice
according to which anomalies exists only when a precise relation exists between number
of derivatives and space--time dimensions. It is also suggested by the presence of
a unique central charge in the underlying $W_\infty$ algebra. 

On the other hand anomalies are not necessary to describe higher moments of the 
Hawking radiation. Rather, the properties of the $W_\infty$ algebra offer
a convincing explanation for them\footnote{After this article was posted 
on the archive, S.Iso and H.Umetsu 
pointed out to us that our result has an additional valence: higher spin
anomalies would give rise to a new kind of 'hairs' corresponding to higher
spin central charges; therefore our proof of the absence of such anomalies 
shows the agreement of the Hawking radiation analysis with 
the no--hair theorem.}. We therefore conclude 
our analysis with the claim that the universal character of the Hawking fluxes 
has its basis in a $W_\infty$ algebra underlying the matter model for radiation.

\acknowledgments

We would like to thank R.Banerjee, S.Iso and H.Umetsu for their
kind and useful messages and P.Dominis Prester for useful discussions. 
M.C. would like to thank SISSA for hospitality and Central European
Initiative (CEI) and The National Foundation for Science, Higher Education
and Technological Development of the Republic of Croatia (NZZ) for
financial support. I.S.,M.C and S.P. would like to acknowledge support 
by the Croatian Ministry of Science, Education and Sport under the 
contract no.119-0982930-1016

\section*{Appendix}
\appendix

\section{The $\tau$ transformations}

In this Appendix we would like to discuss the nature of the $\tau$ 
transformations and argue that they are abelian. Let us start from
the second term in the LHS of \refb{djv4}, a term which does not appear
in the consistent version of the conservation law. The LHS of \refb{djv4}
is formally generated by the variation of the action with respect to 
$\tau$ given by \refb{deltaBtau}  {\it and} by
\be
\delta_\tau g_{\mu\nu} = a \nabla^\l R\,\tau_{\mu\nu\l}\label{deltataug}
\ee
where $a=-\frac {1}5$.

The presence of this nontrivial transformation of the metric under $\tau$
changes completely the rules laid down in section 5.1. Therefore we
must ask ourselves whether \refb{deltataug} is a true symmetry 
transformation or a simple functional variation of the fields necessary in
order to derive a covariant conservation law. We will argue here that
that the second alternative is the correct one. Therefore \refb{deltataug} 
is not a symmetry operation and the rules of section 5.1 are correct,
in particular the $\tau$ transformation rules are abelian. But, let
us explain this in stages.

The way eq.\refb{djv4} was obtained does not allow us to conclude whether
it represents a consistent or covariant conservation law. However the
functional variation \refb{deltataug} contains a non--universal factor
$a$ ($a$ varies according to the regularization and the model) which should
tell us that the latter cannot be a symmetry operation. However, 
short of a conclusive argument, we can try to embed \refb{deltataug}
as well as \refb{deltaBtau} in a new set of transformations, where
possibly $\delta_\tau \tau\neq 0$, and see whether we can implement 
a new group theoretical transformation. This is guaranteed if the
corresponding functional operator $\delta_\tau$ turns out to be nilpotent.
However, as we shall see, this is not the case.

Let us consider a general form for variation $\delta_\tau g_{\mu\nu}$
\begin{equation}
	\delta_\tau g_{\mu\nu} = \sum_{i=1}^{12} a_i I^{i}_{(\mu\nu)}
	\label{eqdeltataug}
\end{equation}
where
\begin{eqnarray}
	I^1_{\mu\nu} &=& \nabla_\mu \nabla_\nu \nabla^\alpha 
\tau_\alpha{}^\beta {}_\beta \CR
	I^2_{\mu\nu} &=& \nabla^\alpha \nabla_\alpha \nabla_\mu 
\tau_\nu{}^\beta {}_\beta \CR
	I^3_{\mu\nu} &=& \nabla^\alpha \nabla^\beta \nabla_\mu 
\tau_\nu{}_\alpha {}_\beta \CR
	I^4_{\mu\nu} &=& \nabla^\alpha \nabla_\alpha \nabla^\beta 
\tau_{\mu\nu}{}_\beta \CR
	I^5_{\mu\nu} &=& \nabla^\alpha \nabla^\beta \nabla^\gamma 
\tau_{\alpha\beta\gamma} g_{\mu\nu} \CR
	I^6_{\mu\nu} &=& \nabla^\alpha \nabla_\alpha \nabla^\beta 
\tau_\beta{}^{\gamma}{}_{\gamma} g_{\mu\nu} \CR
	I^7_{\mu\nu} &=& R \nabla^\alpha \tau_{\mu\nu}{}_\alpha \CR
	I^8_{\mu\nu} &=& R \nabla_\mu \tau_\nu{}^\alpha {}_\alpha \CR
	I^9_{\mu\nu} &=& R \nabla^\alpha 
\tau_\alpha{}^{\beta}{}_{\beta} g_{\mu\nu} \CR
	I^{10}_{\mu\nu} &=& \nabla^\alpha R \tau_{\mu\nu}{}_\alpha \CR
	I^{11}_{\mu\nu} &=& \nabla_\mu    R \tau_\nu{}^\alpha {}_\alpha\CR
	I^{12}_{\mu\nu} &=& \nabla^\alpha R 
\tau_\alpha{}^{\beta}{}_{\beta} g_{\mu\nu} 
	\label{eqdeltataugterms}
\end{eqnarray}
We look at the possible constraints on the coefficients $a_i$ in 
\refb{eqdeltataug} that come from nilpotence of $\delta_\tau$. 
Acting with $\delta_\tau$ on \refb{deltaBtau} we obtain
\begin{eqnarray}
\delta_\tau^2 B^{(4)}_{\mu_1\mu_2\mu_3\mu_4}= 
(\delta_\tau \nabla_{\mu_1}) \tau_{\mu_2\mu_3\mu_4} +
\nabla_{\mu_1} \delta_\tau \tau_{\mu_2\mu_3\mu_4} + { cycl.}
	\label{eqdeltadeltaB}
\end{eqnarray}
The first term gives
\begin{eqnarray}\label{eqdeltadeltaBres}
&&(\delta_\tau \nabla_{\mu_1}) \tau_{\mu_2\mu_3\mu_4} = \CR
&& -6 \left(3 a_3+2 a_{10}\right)  \nabla_{\mu _1}\nabla^{\alpha}R
\tau {}_{\mu _2}{}_{\mu _3}{}^{\beta}\tau{}_{\mu _4}{}_{\alpha}{}_{\beta} 
 -6 \left(3 a_3+2 a_{10}\right)  \nabla^{\alpha}R
\tau {}_{\mu _1}{}_{\mu _2}{}^{\beta}\nabla_{\mu _3}
\tau{}_{\mu _4}{}_{\alpha}{}_{\beta} \CR 
&& +\left(15 a_3+6 a_7\right)   \nabla^{\beta}R
\tau {}_{\mu _1}{}_{\mu _2}{}_{\beta}\nabla^{\alpha}
\tau{}_{\mu _3}{}_{\mu _4}{}_{\alpha}  
 -6 \left(5 a_3+2 a_7\right)   \nabla_{\mu _1}R
\tau {}_{\mu _2}{}_{\mu _3}{}^{\alpha}\nabla^{\beta}
\tau{}_{\mu _4}{}_{\alpha}{}_{\beta} \CR 
&& +\left(9 a_3+6 a_{10}\right)   \nabla^{\alpha}R
\tau {}_{\mu _1}{}_{\mu _2}{}^{\beta}\nabla_{\beta}
\tau{}_{\mu _3}{}_{\mu _4}{}_{\alpha}  
 -6 \left(5 a_3+2 a_7\right)   R
\tau {}_{\mu _1}{}_{\mu _2}{}^{\alpha}\nabla_{\mu _3}
\nabla^{\beta}\tau{}_{\mu _4}{}_{\alpha}{}_{\beta} \CR 
&& +3 a_3   R\tau {}_{\mu _1}{}_{\mu _2}{}_{\mu _3}\nabla^{\alpha}
\nabla^{\beta}
\tau{}_{\mu _4}{}_{\alpha}{}_{\beta}  
 -3 a_3   g{}_{\mu _1}{}_{\mu _2}R
\tau {}_{\mu _3}{}_{\mu _4}{}^{\alpha}\nabla^{\beta}
\nabla^{\gamma}\tau{}_{\alpha}{}_{\beta}{}_{\gamma} \CR 
&& +\left(15 a_3+6 a_7\right)   R
\tau {}_{\mu _1}{}_{\mu _2}{}^{\alpha}\nabla_{\alpha}
\nabla^{\beta}\tau{}_{\mu _3}{}_{\mu _4}{}_{\beta}  
 -6 a_3   \tau {}_{\mu _1}{}_{\mu _2}{}^{\alpha}
\nabla_{\mu _3}\nabla_{\mu _4}\nabla^{\beta}\nabla^{\gamma}
\tau{}_{\alpha}{}_{\beta}{}_{\gamma} \CR 
&& -12 a_4   \tau {}_{\mu _1}{}_{\mu _2}{}^{\alpha}
\nabla_{\mu _3}\nabla^{\beta}\nabla_{\beta}\nabla^{\gamma}
\tau{}_{\mu _4}{}_{\alpha}{}_{\gamma}  
 -12 a_5   \tau {}_{\mu _1}{}_{\mu _2}{}_{\mu _3}\nabla_{\mu _4}
\nabla^{\alpha}\nabla^{\beta}\nabla^{\gamma}
\tau{}_{\alpha}{}_{\beta}{}_{\gamma} \CR 
&& +6 a_4   \tau {}_{\mu _1}{}_{\mu _2}{}^{\alpha}
\nabla_{\alpha}\nabla^{\beta}\nabla_{\beta}\nabla^{\gamma}
\tau{}_{\mu _3}{}_{\mu _4}{}_{\gamma}  
 +6 a_5   g{}_{\mu _1}{}_{\mu _2}\tau {}_{\mu _3}{}_{\mu _4}{}^{\alpha}
\nabla_{\alpha}\nabla^{\beta}\nabla^{\gamma}\nabla^{\delta}
\tau{}_{\beta}{}_{\gamma}{}_{\delta} \CR 
&& +\frac{3}{2} \left(3 a_2-a_3+2 a_8\right)   R^{2}
\tau {}_{\mu _1}{}_{\mu _2}{}_{\mu _3}
\tau{}_{\mu _4}{}^{\alpha}{}_{\alpha}  
 +\left(3 a_3-3 a_2\right)   \nabla_{\mu _1}\nabla^{\beta}R
\tau {}_{\mu _2}{}_{\mu _3}{}_{\beta}\tau{}_{\mu _4}{}^{\alpha}{}_{\alpha} \CR 
&& +3 \left(a_2-a_3\right)   \nabla^{\beta}\nabla_{\mu _1}R
\tau {}_{\mu _2}{}_{\mu _3}{}_{\beta}\tau{}_{\mu _4}{}^{\alpha}{}_{\alpha}  
 +6 \left(a_2+a_3-2 a_{12}\right)   \nabla_{\mu _1}\nabla^{\alpha}R
\tau {}_{\mu _2}{}_{\mu _3}{}_{\mu _4}\tau{}_{\alpha}{}^{\beta}{}_{\beta} \CR 
&& +\frac{3}{2} \left(-3 a_2+a_3-2 a_8\right)   g{}_{\mu _1}{}_{\mu _2}
R^{2}\tau {}_{\mu _3}{}_{\mu _4}{}^{\alpha}\tau{}_{\alpha}{}^{\beta}{}_{\beta}  
 -3 \left(a_2-a_3+2 a_{11}\right)   \nabla_{\mu _1}\nabla_{\mu _2}R
\tau {}_{\mu _3}{}_{\mu _4}{}^{\alpha}
\tau{}_{\alpha}{}^{\beta}{}_{\beta} \CR 
&& -3 \left(a_2+a_3-2 a_{12}\right)   g{}_{\mu _1}{}_{\mu _2}
\nabla^{\alpha}\nabla^{\beta}R\tau {}_{\mu _3}{}_{\mu _4}{}_{\alpha}
\tau{}_{\beta}{}^{\gamma}{}_{\gamma}  
 +6 \left(a_2+a_8-a_{11}\right)   \nabla^{\beta}R
\tau {}_{\mu _1}{}_{\mu _2}{}_{\beta}
\nabla_{\mu _3}\tau{}_{\mu _4}{}^{\alpha}{}_{\alpha} \CR 
&& +6 \left(a_2+a_3-2 a_{12}\right)   \nabla^{\alpha}R
\tau {}_{\mu _1}{}_{\mu _2}{}_{\mu _3}\nabla_{\mu _4}
\tau{}_{\alpha}{}^{\beta}{}_{\beta}  
 -6 \left(2 a_2-a_3+a_8+a_{11}\right)   \nabla_{\mu _1}R
\tau {}_{\mu _2}{}_{\mu _3}{}^{\alpha}
\nabla_{\mu _4}\tau{}_{\alpha}{}^{\beta}{}_{\beta} \CR 
&& +12 \left(a_2+a_3-a_9\right)   \nabla_{\mu _1}R
\tau {}_{\mu _2}{}_{\mu _3}{}_{\mu _4}\nabla^{\alpha}
\tau{}_{\alpha}{}^{\beta}{}_{\beta}  
 -6 \left(a_2+a_3-a_9\right)   g{}_{\mu _1}{}_{\mu _2}
\nabla^{\gamma}R\tau {}_{\mu _3}{}_{\mu _4}{}_{\gamma}\nabla^{\alpha}
\tau{}_{\alpha}{}^{\beta}{}_{\beta} \CR 
&& -6 \left(a_2+a_8-a_{11}\right)   \nabla_{\mu _1}R
\tau {}_{\mu _2}{}_{\mu _3}{}^{\alpha}\nabla_{\alpha}
\tau{}_{\mu _4}{}^{\beta}{}_{\beta}  
 -3 \left(a_2+a_3-2 a_{12}\right)   g{}_{\mu _1}{}_{\mu _2}
\nabla^{\alpha}R\tau {}_{\mu _3}{}_{\mu _4}{}^{\beta}\nabla_{\beta}
\tau{}_{\alpha}{}^{\gamma}{}_{\gamma} \CR 
&& +3 \left(-3 a_2+a_3-2 a_8\right)   R
\tau {}_{\mu _1}{}_{\mu _2}{}^{\alpha}\nabla_{\mu _3}
\nabla_{\mu _4}\tau{}_{\alpha}{}^{\beta}{}_{\beta}  
 +6 a_6   g{}_{\mu _1}{}_{\mu _2}\tau {}_{\mu _3}{}_{\mu _4}{}^{\alpha}
\nabla_{\alpha}\nabla^{\beta}\nabla_{\beta}\nabla^{\gamma}
\tau{}_{\gamma}{}^{\delta}{}_{\delta} \CR 
&& +3 a_2   R\tau {}_{\mu _1}{}_{\mu _2}{}_{\mu _3}\nabla^{\alpha}
\nabla_{\alpha}\tau{}_{\mu _4}{}^{\beta}{}_{\beta}  
 -3 a_2   g{}_{\mu _1}{}_{\mu _2}R\tau {}_{\mu _3}{}_{\mu _4}{}^{\alpha}
\nabla^{\beta}\nabla_{\beta}\tau{}_{\alpha}{}^{\gamma}{}_{\gamma} \CR 
&& -3 \left(a_1+2 \left(a_2+a_3-a_9\right)\right)   
g{}_{\mu _1}{}_{\mu _2}R\tau {}_{\mu _3}{}_{\mu _4}{}^{\alpha}
\nabla_{\alpha}\nabla^{\beta}\tau{}_{\beta}{}^{\gamma}{}_{\gamma}  
 -6 a_2   \tau {}_{\mu _1}{}_{\mu _2}{}^{\alpha}\nabla_{\mu _3}
\nabla_{\mu _4}\nabla^{\beta}\nabla_{\beta}\tau{}_{\alpha}{}^{\gamma}{}_{\gamma} \CR 
&& -6 a_1   \tau {}_{\mu _1}{}_{\mu _2}{}^{\alpha}\nabla_{\mu _3}
\nabla_{\mu _4}\nabla_{\alpha}\nabla^{\beta}\tau{}_{\beta}{}^{\gamma}{}_{\gamma}  
 -12 a_6   \tau {}_{\mu _1}{}_{\mu _2}{}_{\mu _3}\nabla_{\mu _4}
\nabla^{\alpha}\nabla_{\alpha}\nabla^{\beta}\tau{}_{\beta}{}^{\gamma}{}_{\gamma} \CR 
&&
 +3 \left(a_1+4 \left(a_2+a_3-a_9\right)\right)   R
\tau {}_{\mu _1}{}_{\mu _2}{}_{\mu _3}\nabla_{\mu _4}\nabla^{\alpha}
\tau{}_{\alpha}{}^{\beta}{}_{\beta} \CR 
\end{eqnarray}
with the symmetrization understood on both sides of the equation.
These terms need to be canceled by the RHS in
\refb{eqdeltadeltaB}
\begin{equation}
\nabla_{\mu_1} \delta_\tau \tau_{\mu_2\mu_3\mu_4}
	\label{eqdeltadeltaB2}
\end{equation}
It follows that the coefficients in front of all the terms in 
\refb{eqdeltadeltaBres} which do not contain any $\nabla_{\mu_i}$ 
must be zero. This gives a system of equations for the coefficients $a_i$
with a solution that all $a_i$ are zero except $a_{11}$. 
Now we argue that, also, $a_{11}$ must be zero.
The variation $\delta_\tau$ of metric reads
\begin{equation}
	\delta_\tau g_{\mu\nu} = a_{11}	\nabla_\mu   
 R \tau_\nu{}^\alpha {}_\alpha 
		\label{eqdeltataugterms2}
\end{equation}
Then, $\delta_\tau^2 B_{\mu_1\mu_2\mu_3\mu_4}$ is reduced to
\begin{eqnarray}
	\label{eqdeltadeltaBres2}
\delta_\tau^2 B^{(4)}_{\mu_1\mu_2\mu_3\mu_4}=4 \nabla_{\mu_1} 
\delta_\tau \tau_{\mu_2\mu_3\mu_4}
& -24& a_{11}  \nabla_{\mu _1}\nabla_{\mu _2}R
\tau {}_{\mu _3}{}_{\mu _4}{}^{\alpha}\tau{}_{\alpha}{}^{\beta}{}_{\beta}
\CR 
& -24& a_{11}  \nabla^{\beta}R\tau {}_{\mu _1}{}_{\mu _2}{}_{\beta}
\nabla_{\mu _3}\tau{}_{\mu _4}{}^{\alpha}{}_{\alpha} \CR 
& -24& a_{11}  \nabla_{\mu _1}R\tau {}_{\mu _2}{}_{\mu _3}{}^{\alpha}
\nabla_{\mu _4}\tau{}_{\alpha}{}^{\beta}{}_{\beta} \CR 
& +24& a_{11}  \nabla_{\mu _1}R\tau {}_{\mu _2}{}_{\mu _3}{}^{\alpha}
\nabla_{\alpha}\tau{}_{\mu _4}{}^{\beta}{}_{\beta}  
\end{eqnarray}
where symmetrization in $\mu_i$ is understood on the right hand side. 
Note that in each term in \refb{eqdeltadeltaBres2} there is a $\tau$ 
with two external indices without any derivative acting on it. So, we 
conclude that $a_{11}$ must be zero because
whatever choice we take for $ \delta_\tau \tau_{\mu_2\mu_3\mu_4}$, the 
term $\nabla_{\mu_1} \delta_\tau \tau_{\mu_2\mu_3\mu_4}$ 
will not be able to cancel the last four terms in \refb{eqdeltadeltaBres2}. 

In summary, we have shown that 
\begin{equation}
	\delta_\tau g_{\mu\nu}  = 0
	\label{eqdeltataugzero}
\end{equation}
no matter what $\delta_\tau \tau_{\mu_2\mu_3\mu_4}$ is. 
In turn, eq.~\refb{eqdeltataugzero}, together with 
eq.~\refb{eqdeltadeltaB}, implies that 
\begin{equation}
	\delta_\tau \tau_{\mu \nu \rho} = 0
	\label{eqdeltatautauzero}
\end{equation}

To conclude this Appendix, let us justify the claim we made that 
eq.\refb{Deltaomega} is the most general $2n$--th order covariant cocycle.
To this end we prove that terms of the form
\be
\omega^{\mu_{1} \dots \mu_{2n-1}} \, \epsilon_{\mu_{1} \alpha_{1}} \, 
\epsilon_{\mu_{2} \alpha_{2}} \dots \epsilon_{\mu_{k} \alpha_{k}} \, 
\nabla^{\alpha_{1}} \nabla^{\alpha_{2}} \dots \nabla^{\alpha_{k}} 
\nabla_{\mu_{k+1}} \dots \nabla_{\mu_{2n-1}} R \label{epsilonepsilon}
\ee
are in fact \emph{equivalent} to either $I^{\omega}$ or $K^{\omega}$. 
Using the formula valid in $D=2$,
\be
\epsilon_{\alpha \beta} \epsilon_{\mu \nu} = 
g_{\alpha \mu} g_{\beta \nu} - g_{\alpha \nu} g_{\beta \mu}\label{epsilong}
\ee
we can eliminate the $\epsilon$-tensors two by two in \refb{epsilonepsilon}. 
In every step we produce two terms, out of which the first is zero because of 
the tracelessness of $\omega$ and the second contracts two indices 
of $\omega$ with the indices of the covariant derivatives on the right side. 
The form of the final expression will depend on parity of $k$; in the case of 
even $k$ we get $(-1)^{k/2} \, I^{\omega}$ and in the case of odd $k$, 
$(-1)^{(k-1)/2} \, K^{\omega}$. Therefore, all such terms are already 
included in the general form of $\Delta_{\omega}$, eq.\refb{Deltaomega}.

\section{Diff cocycles}

This Appendix is devoted to eqs.\refb{deltaxi2} and \refb{deltaxi4}.
Our result is that while \refb{deltaxi2} has a nontrivial solution,
the solutions to eq.\refb{deltaxi4} are all trivial.
The solution to \refb{deltaxi2} is the well--known diffeomeorphim 
anomaly in 2d, therefore we will concentrate on \refb{deltaxi4}.
Although the same method could be easily applied to find the explicit 
solutions to  \refb{deltaxi2}, we will not do it here. In the sequel
of this Appendix the terms covariance and covariant are used
with reference to diffeomorphisms. 

The basic idea of this Appendix is to apply the results of \cite{BPT}
by adapting them to the present case. First of all let us notice that
the anomaly analysis is carried out in the Euclidean. We will denote
Euclidean tensor indices by lower case Latin letters.

Let us start from a general result in \cite{BPT}. The general form of the
solutions to \refb{deltaxi4} is
\be
\Delta^{(4)}_\xi = \int d^2x \, \left(\d_m \xi^m \, {\mathfrak b}
+ \d_{p_1} \d_{p_2} \xi^m {\mathfrak b}_m^{p_1 p_2} \right)\label{gensol}
\ee
where ${\mathfrak b}$ and ${\mathfrak b}_{p_1p_2}^m$ are polynomial 
expressions 
of the fields and their derivatives in which all the indices are saturated 
except for the explicitly shown ones, and ${\mathfrak b}$ is not itself a 
derivative. Notice that ${\mathfrak b},{\mathfrak b}_{p_1p_2}^m$ 
are {\it not}, in general, covariant tensors. For future
reference let us call {\it first type} and {\it second type}
the cocycles having the form of the first and second term in the RHS
of \refb{gensol}, respectively.

We stress that in this appendix we cannot, in general, use covariance 
as a classifying
device. This obliges us to trudge our way through a multitude
of inelegant and unfamiliar formulas.
 
The first type cocycles were discussed both in \cite{Bonora84} 
and \cite{Bonora85}.
Any such cocycle\footnote{The corresponding anomaly does not contain
the $\epsilon_{\mu\nu}$ tensor with an unsaturated index, see eq.\refb{diffanom}.} 
is a partner of a Weyl cocycle and can be eliminated
in favor of the partner by subtracting a suitable counterterm. Since
we have shown in \cite{BC} that, at order four, there are no non--trivial
Weyl cocycles (trace anomalies), we will disregard these cocycles
altogether and concentrate on cocycles of the second type in \refb{gensol},
i.e. on cocycles proportional to $\d_{p_1} \d_{p_2} \xi^m$.  
It is easy to realize that ${\mathfrak b}_{p_1p_2}^m$ can be 
synthetically written in the following general form
\be
{\mathfrak b}= A_1 + \Gamma A_2 +\Gamma \Gamma A_3+
\Gamma \Gamma \Gamma A_4+\d \Gamma A_5 +\Gamma \d \Gamma A_6+ 
\d\d \Gamma A_7\label{genformb}
\ee
where we have understood all the indices (for instance $A_1$ stands for 
${A_1}_{m}^{p_1p_2}$) and $A_1, \ldots A_7$ are for weight 1 
covariant tensors.
The symbol $\Gamma$ represents the linear (not necessarily metric)
connection $\Gamma_{nm}^l$.

An important remark is that, since $\Delta^{(4)}_\xi$ is degree 4,
it follows that all the expressions $A_i$ can only be linear in the 
background field $B^{(4)}$ and contain $4-i$ derivatives
for $i=1,2,3,4$, one derivative for $i=5$ and no derivatives 
for $i=6,7$.  

As we said before the fact that we cannot use covariance in expressing
${\mathfrak b}_{p_1p_2}^m$ is a tremendous complication. There are
however expedients one can use to simplify one's life. One such
contrivance consists in splitting the functional operator $\delta_\xi$
into two parts
\be
\delta_\xi = \delta^c_\xi+ \hat\delta_\xi\label{splitdeltaxi}
\ee
where $\delta^c_\xi$ acts on cochains as if they where covariant tensors,
while $\hat\delta_\xi$ represents the noncovariant part  
of the $\delta_\xi$ action. 

For instance we have
\be
&& \hat\delta_\xi \,\Gamma_{mn}^l= \d_m\d_n\xi^l\0\\
&&\hat \delta_\xi \, \d_k \Gamma_{mn}^l= \d_k\d_n\d_m \xi^l +\d_k\d_m\xi^p \Gamma_{pn}^l
+\d_k\d_n\xi^p \Gamma_{pm}^l- \d_k\d_p\xi^l \Gamma_{mn}^p\0\\
&& \hat\delta_\xi \,\xi^l =- \xi^n\d_n\xi^l\0\\
&& \hat\delta_\xi\, \d_n \xi^l= - \d_n\xi^m\d_m\xi^l\0\\
&&  \hat\delta_\xi\, \d_m \d_n \xi^l=0\0\\
&&  \hat\delta_\xi\, \d_m \d_n \d_p\xi^l= \d_m \d_n \xi^q \,\d_q\d_p\xi^l+
\d_p \d_m \xi^q \,\d_q\d_n\xi^l+\d_n \d_p \xi^q \,\d_q\d_m\xi^l
\label{hatdelta}
\ee
It is easy to prove that 
\be
 \hat\delta_\xi^2=0,\0
\ee
but one must be careful: in general $\hat\delta_\xi$ does not commute with 
the operation of differentiation except when particular conditions are met.
The latter include the cases when $\hat\delta_\xi$ acts on forms or 
on expressions without unsaturated indices.

It is convenient to write the integrand of \refb{gensol} as a two--form.
So we write
\be
\Delta_\xi=\int d^2x \, \d_{p_1} \d_{p_2} \xi^m {\mathfrak b}_m^{p_1 p_2}
\equiv\int Q_2^1\label{Q21}
\ee
The lower index in $Q_n^i$ represents the form order, the upper index 
denotes the ghost number (number of $\xi$ factors).
We must have $\delta_\xi \Delta_\xi =\hat \delta_\xi \Delta_\xi=0$. Therefore
\be
\hat \delta_\xi Q_2^1 = d Q_1^2 \label{Q12}
\ee
for some one--form $Q_1^2$. Applying $\hat \delta_\xi$ to both sides
of this equation and the local Poincar\'e lemma\footnote{By local 
Poincar\'e lemma we mean a basic property of local field theory:
if a p--form is a polynomial made of local fields and their derivatives,
whose exterior derivative vanishes, either it is a top form, or it is a
constant if it is a 0--form, or it is a total derivative. This is an
off--shell statement.}, we get
\be
\hat \delta_\xi Q_1^2= d Q_0^3\label{Q03}
\ee
and, of course, $\hat \delta_\xi Q_0^3=0$. The reason why we introduce
these descent equations is that the classification problem is easier
on lower order forms (with higher ghost number) than on higher
order forms. Briefly stated the strategy consist in chopping off 
as many coboundaries and first type cocycles as possible, so as to be
left with a subset of possibilities which can be easily dealt with. 

Schematically, first one proves that solutions to
$\hat \delta_\xi (Q_2^1-dP_1^1)=0$, where $Q_2^1$ is defined by \refb{Q21},
either correspond to first type 
cocycles or are trivial. As a consequence of this one proves that solutions to
$\hat \delta_\xi (Q_1^2-dP_0^2)=0$, where $Q_1^2$ is defined by 
\refb{Q12}, are trivial. Thus possible non-trivial second type cocycles 
are to be looked for among the $Q_0^3$ (defined by \refb{Q03}) that do not 
vanish (up to a diff transformation), of which none exist.

Let us go now to a more detailed description. We need to introduce some 
notation. Let  $\omega$ be a 0,1 or 2--form with component $\omega,\omega_m,
\omega_{nm}$, respectively. We define the dual tensor
\be
\tilde \omega^{nm} = \varepsilon^{nm}\omega,\
\tilde \omega^{n} = \varepsilon^{nm}\omega_m,\quad\quad
\tilde \omega=  \varepsilon^{nm}\omega_{nm}\0
\ee
where $\varepsilon$ is the constant antisymmetric symbol. We remark
that if $\omega$ is an exact 1--form, the corresponding dual tensor
is a divergence. Next let us introduce a distinction which is basic in the
economy of our proof: we separate all the cochains
$Q_n^i, P_n^i$ into two classes, class A and class B.
Any term is class A if it contains only $\d\d\xi$ or higher derivatives of 
$\xi$, it is class B otherwise. 

We are now ready to state the first lemma.

{\it Lemma 1.} A cocycle $\Delta_\xi$ in \refb{Q21} that satisfies
\be
\hat\delta_\xi (Q_2^1-dP_1^1)=0\label{lemma1}
\ee
is either a first type cocycle or a coboundary. In the latter case
$P_1^1$ can be chosen to be class A.

The proof in \cite{BPT} (Theorem 5.1 there) applies to the
present case with obvious modifications. Let us give just one example 
(out of many).
The expression $\Omega_\xi=\sqrt{-g} \d \d\xi B \nabla R$ 
($B\equiv B^{(4)}$), with indices 
contracted in all possible ways, is an example of \refb{lemma1} where 
$Q_2^1$ is given by the dual of $\Omega_\xi$ and $P_1^1$ vanishes. 
It corresponds in fact to a coboundary generated by the counterterm
$\int \sqrt{-g}\Gamma B \nabla R$ where the $\Gamma$ indices are
contracted in the same way as the $\d\d\xi$ indices in $\Omega_\xi$.

From Lemma 1 and (\ref{Q21},\ref{Q12}), it is easy to show that $Q_1^2$ 
corresponds to a coboundary $Q_2^1$ if and only if
\be
Q_1^2=\hat\delta_\xi P_1^1 +dP_0^2\label{Q12P}
\ee
for some class A (or vanishing) $P_1^1$ and some ghost number 2 0--form 
$P_0^2$.

Another piece of independent information is provided by the following

{\it Lemma 2.} $Q_1^2$ defined by eq.\refb{Q12} can be written in a class
A form, that is in a form bilinear either in $\d\d\xi$ or linear in both
$\d\d\xi$ and $\d\d\d\xi$.

The proof is as follows. The dual tensor to $Q_1^2$ can be written in the
general form
\be
&&
\xi\xi F_1 + \xi\d\xi F_2+\xi\d\d\xi F_3 + \xi\d\d\d\xi F_4+
\xi\d\d\d\d\xi F_5 + \xi\d\d\d\d\d\xi F_6+ \d\xi\d\xi F_7\0\\
&&\quad\quad +
\d\xi\d\d\xi F_8 + \d\xi\d\d\d\xi F_9 + \d\xi\d\d\d\d\xi F_{10}+
 \d\d\xi\d\d\xi F_{11}+\d\d\xi\d\d\d\xi F_{12} \label{xixiF}
\ee
where $F_i$ with $i=1,\ldots,12$ are in general not tensors: they may 
contain $\Gamma$ factors. For simplicity 
all the indices are understood. For instance $\xi\xi F_1$ stands
for $\xi^i\xi^j {F_1}_{ij}{}^l$. 

Now one can see that, as a consequence 
of (\ref{genformb}, \ref{hatdelta}) and \refb{Q21}, $dQ_1^2$ must
be class A. Then, dualizing, we see that, in particular,
it must be $\d_l {F_1}_{ij}{}^l=0$. This implies, by the local Poincar\'e 
lemma, that ${F_1}_{ij}{}^l= \d_m {F}_{ij}{}^{lm}$ for a suitable 
tensor $F$ antisymmetric in $l,m$. But then we can write
\be
\d_l( \xi^i\xi^j {F_1}_{ij}{}^l) = \d_m ((\d_l\xi^i \xi^j+\xi^i\d_l\xi^j)
F_{ij}{}^{lm})\0
\ee
This means that $F_1$ can be absorbed into $F_2$. We can repeat the same
trick on the other terms. $F_6$ and $F_{10}$ do not contain derivatives,
therefore they must vanish. All the other terms can be reduced to the form 
$F_{11}$ and $F_{12}$. This proves the lemma.

It follows from \refb{Q12P} that $dP_0^2$ is class A. But then, using the 
same argument as in the previous lemma, it is easy to see that $P_0^2$
itself is class A.

Next one can prove:

{\it Lemma 3.} $Q_1^2$, given by eq.\refb{Q12}, is a coboundary
if and only if
\be
\hat\delta_\xi (Q_1^2-dP_0^2)=0\label{lemma3}
\ee
where $P_0^2$ is class A or 0. 

The {\it only if} part follows immediately
by applying $\hat\delta_\xi$ to \refb{Q12P}  and using the previous remark.
The proof of the {\it if} part is more complicated. We give it here in some detail.
We have to prove that if $Q_1^2$ satisfies (\ref{lemma3}) then it can be 
written in the form \refb{Q12P}. Let us write down
the general form of the components of $Q_1^2$:
\be
Q_1^2: \quad \d\d\xi\,\d\d\xi(A_l + \Gamma B_l)+ \d\d\xi\,\d\d\d\xi C_l\label{gfQ12}
\ee
where $A_l,B_l,C_l$ are covariant tensors. All the indices have been understood except
the one--form component index $l$. It it easy to see that $dP_0^2$, being class A,
can be absorbed into $Q_1^2$; therefore we will not indicate it explicitly. The tensors
in \refb{gfQ12} have evident symmetry properties in the indices which we will not
spell out. Acting now with $\hat\delta_\xi$ on $Q_1^2$, after
some simple algebra we get:
\be
\d_{p_1}\d_{p_2}\xi^i \d_{q_1}\d_{q_2}\xi^j \left( \d_{r_1}\d_{r_2}\xi^k \,
B_{lijk}^{p_1p_2q_1q_2r_1r_2}-3 \d_{j}\d_{q_3}\xi^k\, C_{lik}^{p_1p_2q_1q_2q_3}\right)\0
\ee
To satisfy \refb{lemma3} this must vanish, which implies the constraint
\be
B_{lijk}^{p_1p_2q_1q_2r_1r_2}=  3 C_{lik}^{p_1p_2q_1q_2r_1}\delta_j^{r_2} \label{B3C}
\ee
while the tensor $A_l$ is unconstrained.

Now let us write a class A $P_1^1$ in the form 
\be
P_1^1:\quad \d\d\xi(\Gamma\, K_l + \Gamma \,\Gamma\, L_l + \d \Gamma\, M_l)\label{P11}
\ee 
with the same conventions as above. There are other possible terms one could add, but these
will suffice. After operating with $\hat \delta_\xi$ we get
\be
\hat \delta_\xi P_1^1:\quad &&-\d_{p_1}\d_{p_2}\xi^i \left[ \d_{q_1}\d_{q_2}\xi^j\,
K_{lij}^{p_1p_2q_1q_2}+ \d_r \d_{q_1}\d_{q_2}\xi^j\,M_{lij}^{p_1p_2q_1q_2r}\right. 
\label{deltaxiP11}\\
&& + \left. \d_{q_1}\d_{q_2}\xi^j\,\Gamma_{r_1r_2}^k \left( L_{lijk}^{p_1p_2q_1q_2r_1r_2}+
2 M_{lik}^{p_1p_2q_1r_1q_2}\delta_j^{r_2} -M_{lij}^{p_1p_2r_1r_2q_1}\delta_k^{q_2}\right)
\right]\0
\ee
From this we see that $\hat \delta_\xi P_1^1$ reproduces $Q_1^2$ provided
\be
&&K_{lij}^{p_1p_2q_1q_2}=-A_{lij}^{p_1p_2q_1q_2}\0\\
&&M_{lij}^{p_1p_2q_1q_2q_3}=-C_{lij}^{p_1p_2q_1q_2q_3}\0\\
&&L_{lijk}^{p_1p_2q_1q_2r_1r_2}= C_{lik}^{p_1p_2q_1q_2(r_1}\delta_j^{r_2)}+
C_{lij}^{p_1p_2r_1r_2(q_1}\delta_k^{q_2)}\label{KLM}
\ee
where indices in round brackets are meant to be symmetrized. We notice
that this implies in particular that the tensor $M$ must be chosen symmetric
in the $q_1,q_2,q_3$ indices. This proves Lemma 3.

At this point our quest comes to an end, because from \refb{lemma3} and 
\refb{Q03}, it follows that
\be
d(Q_0^3-\hat \delta_\xi P_0^2)=0\label{P02}
\ee
Therefore, since we want to find possible non--trivial second type 
cocycles, we have to look among the $Q_0^3$ that do not satisfy
eq.\refb{P02}. We will prove that there exist none. One way to see
it is as follows. Since $Q_1^2$ is class A (lemma 2), also 
$\hat \delta_\xi Q_1^2$ is. Using an argument similar to
the one which leads to Lemma 2, from \refb{Q03} one can show that 
$Q_0^3$ can be written in the form
\be
\d_{p_1} \d_{p_2} \xi^i \, \d_{q_1} \d_{q_2} \xi^j \, \d_r\xi^k \,
E^{p_1p_2q_1q_2 r}_{ijk}  
\label{EeG}
\ee
where $E$ is a tensor  
antisymmetric in the exchange of the triple $\{p_1,p_2,i\}$
with $\{q_1,q_2,j\}$. 
 
Next, differentiating the $E$ term we get
\be
&&2\d_s\d_{p_1} \d_{p_2} \xi^i \, \d_{q_1} \d_{q_2} \xi^j \, \d_r\xi^k \,
E^{p_1p_2q_1q_2 r}_{ijk} +
\d_{p_1} \d_{p_2} \xi^i \, \d_{q_1} \d_{q_2} \xi^j \,\d_s\d_r\xi^k
 E^{p_1p_2q_1q_2 r}_{ijk}\0\\
&&\quad\quad+
\d_{p_1} \d_{p_2} \xi^i \,  \d_{q_1} \d_{q_2} \xi^j \, \d_r\xi^k
 \d_sE^{p_1p_2q_1q_2 r}_{ijk}\0
\ee
Since $dQ_0^3$ must be class A, we must have in particular $\d_sE=0$.
However $E$ is a tensor linear in the $B^{(4)}$ tensor and does not
contain any 
derivatives. It is evident that $\d_sE=0$ cannot be satisfied, unless
$E\equiv 0$.

In conclusion there are no nontrivial solutions to \refb{deltaxi4}.

\end{document}